\documentclass[preprint,aps,pdflatex,onecolumn]{revtex4} 
\usepackage{amsmath}
\usepackage{amsfonts}
\usepackage{amssymb}
\usepackage{bm}
\usepackage{graphicx}
\usepackage{float}
\usepackage{dcolumn}
\usepackage{multirow}

\newcommand{\sech}{\mathrm{sech}}
\newcommand{\be}{\begin{equation}}
\newcommand{\ee}{\end{equation}}
\newcommand{\bes}{\begin{subequations}}
\newcommand{\ees}{\end{subequations}}
\newcommand{\ben}{\begin{eqnarray}}
\newcommand{\een}{\end{eqnarray}}

\begin{document}
\title{Kink scattering in a hybrid model }
 \author{D. Bazeia${^1}$, Adalto R. Gomes$^{2}$, K. Z. Nobrega$^{3}$, Fabiano C. Simas$^{4}$, }
 \email{bazeia@fisica.ufpb.br, argomes.ufma@gmail.com, bzuza1@yahoo.com.br, simasfc@gmail.com}
 \noaffiliation
\affiliation{
$^1$ Departamento de F\'isica, Universidade Federal da Para\'iba, 58051-970, Jo\~ao Pessoa, PB, Brazil\\
$^2$ Departamento de F\'isica, Universidade Federal do Maranh\~ao (UFMA)
Campus Universit\'ario do Bacanga, 65085-580, S\~ao Lu\'\i s, Maranh\~ao, Brazil\\
$^3$ Departamento de Eletro-Eletr\^onica, Instituto Federal de Educa\c c\~ao, Ci\^encia e
Tecnologia do Maranh\~ao (IFMA), Campus Monte Castelo, 65030-005, S\~ao Lu\'is, Maranh\~ao, Brazil\\
$^4$ Centro de Ci\^encias Agr\'arias e Ambientais-CCAA, Universidade Federal do Maranh\~ao
(UFMA), 65500-000, Chapadinha, Maranh\~ao, Brazil
}
\noaffiliation

\begin{abstract}
In this work we consider a model where the potential has two topological sectors connecting three adjacent minima, as occurs with the $\phi^6$ model. In each topological sector, the potential is symmetric around the local maximum.  For $\phi>0$ there is a linear map between the model and the $\lambda\phi^4$ model. For $\phi<0$ the potential is reflected.  Linear stability analysis of kink and antikink lead to discrete and continuum modes related by a linear coordinate transformation to those known analytically for the $\lambda\phi^4$ model. Fixing one topological sector, the structure of antikink-kink scattering is related to the observed in the $\lambda\phi^4$ model. For kink-antikink collisions a new structure of bounce windows appear. Depending on the initial velocity, one can have oscillations of the scalar field at the center of mass even for one bounce, or a change of topological sector. We also found a structure of one-bounce, with secondary windows corresponding to the changing of the topological sector accumulating close to each one-bounce windows. The kink-kink collisions are characterized by a repulsive interaction and there is no possibility of forming a bound state.

\end{abstract}

\keywords{kink, lower dimensional models, extended classical solutions}

\maketitle


\section{ Introduction }

Spatially localized topological configurations in nonlinear field theories are solutions with localized energy density that attain topological profile and propagate freely in time without loosing form. 
Solitons, for instance, maintain their form even after scattering, but there are other localized structures that present different but still interesting features when they collide with one another. 
On general grounds, the simplest localized solutions in field theories that present topological profile are kinks and antikinks in (1, 1) spacetime dimensions and can be constructed in theories with one or more scalar fields.

Orbifold brane collisions were considered in cosmology, in cyclic/ekpyrotic models of the universe \cite{cyclic1,cyclic2}. Such models are alternatives to the inflationary universe, where the big bang is a transition from a phase of contraction to a phase of expansion.  Local nongaussianities predicted by cyclic models are constrained by the CMB temperature and E-mode polarization maps \cite{planck}. Despite originally formulated in heterotic M-theory, some of the ideas of cyclic models can be considered in other brane universe theories \cite{cyclic1}, such as brane models constructed with kinks embedded in spacetimes with larger dimensions.  

Kink scattering in integrable systems is surprisingly simple, with the solitons gaining at most a phase shift. Some examples of integrable models are: i) KdV equation, connected to the Fermi-Pasta-Ulan problem \cite{fpu1,fpu2} in the continuum limit; ii) nonlinear Schr\"odinger equation, important for describing nonlinear effects in fiber optics \cite{fo1,fo2}); iii) the ubiquitous sine-Gordon equation \cite{sg}, studied among other things in theories describing DNA \cite{dna} and Josephson juntions \cite{jj1,jj2,jj3}.

In nonintegrable models, kink scattering has a complex behavior. For ultrarelativistic velocities and arbitrary potentials, there is an analytical expression for the phase shift \cite{clash}. The simplest nonintegrable and largely studied is the $\lambda\phi^4$ model \cite{sug,moshir,csw,w1,campbell,belkud,aom,gh,saada,doreyrom}. In that model, for larger initial velocities $v$ we have inelastic scattering, with the pair of solitons colliding once and separating thereafter. For smaller velocities than a critical one, $v<v_c$, the kink-antikink forms a composed state named bion that radiates continuously until the complete anihilation of the pair. For smaller velocities with $v\lesssim v_{crit}$ there are regions in velocity, named two-bounce windows, where the scalar field at the center of mass bounces twice before the final separation of the pair. Stability analysis of the $\lambda\phi^4$ kink leads to a Schr\"odinger-like equation with two discrete eigenstates: a zero or translational mode, related to the translational invariance of the model and a vibrational mode.

An argument for the occurrence of two-bounce windows was presented by Campbell, Schonfeld and Wingate (CSW) \cite{csw} as a resonance mechanism for the exchange of energy between the translational and the vibrational mode. A counter-example of this mechanism was found for the asymmetric $\phi^6$ model, where the presence of two-bounce windows is explained not by the (absent) vibrational state of the kink, but due to the presence of vibrational state related to the combined kink-antikink configuration \cite{domerosh}. Another counter example is the total suppression of two-bounce windows due to the presence of multiple vibrational states \cite{sgno}.

Kink scattering in nonintegrable models is a topic that has been under intense investigation. One can cite studies with polynomial models with one \cite{domerosh,dedeke,gankudli,weigel,roman,begani1,galeli,begani2} and two \cite{harosh,alonso1,alonso2,vakiru} scalar fields, nonpolynomial models  \cite{peycam,gankud,gaaes,bazbegani1,bazbegani2,sgn}, vector solitons \cite{yatan},  multi-kinks \cite{mgsdj,masd,sadmke,amdskzsd}, interaction with a boundary \cite{dhmr, adp}, models with generalized dynamics \cite{twin} and interactions with impurities \cite{gooha,feikiva,goohol}.

Recently, using molecular dynamics simulations, it was proposed the realization of nonintegrable $\lambda\phi^4$ kinks in buckled graphene nanoribbon \cite{graph1, graph2}. Kinks in such a system could be a possibility of experimental verification of negative pressure radiation effects \cite{npress1,npress2}. Ab-initio excited-state dynamics was used for study the photogeneration and time evolution of topological excitations in {\it trans}-polyacetilene. Upon lattice relaxation, the produced pair exhibits a pattern of two-bounce resonance characteristic of nonintegrable dynamics \cite{poly}. Then, the investigation of models leading to new patterns of kink scattering in nonintegrable theories could be useful not only to better understand some aspects of nonlinearity, but also to interpret scattering results after ab initio calculations describing these or similar physical systems.

In this work we investigate kink scattering in a hybrid model with two topological sectors. Each  topological sector connects adjacent vacua of the theory. The hybrid model appeared very recently in the Ref.~\cite{bazbem}, generated through the reconstruction of field theories from reflectionless symmetric quantum mechanical potentials. In that paper one has the construction of two distinct field theories from the very same quantum mechanical potential. This motivated us to further explore this model, with the aim to identify its specific features under kink collisions. We notice that the presence of the two sectors seems to simulate the $\phi^6$ model studied in the Ref. \cite{domerosh}. On the other hand, the model can be described as patched from two parts of the $\lambda\phi^4$ potential. However, the model is still different from the $\lambda\phi^4$
model, because of the reflection symmetry. Then a last motivation for considering the hybrid model is to analyze which modification the reflection symmetry introduces to the known
classic results of the $\lambda\phi^4$ model.

Fixing one topological sector, the numerical analysis of antikink-kink scattering shows the expected structure of one-bounce, bion and two-bounce windows explained by the CSW mechanism for a kink (and antikink) with  translational and one vibrational state. However, the analysis of kink-antikink scattering gives unexpected results, such as one-bounce windows and thinner one-bounce windows related to the change of topological sector. We explain our findings of one-bounce windows with the CSW mechanism, adapted for the one-bounce windows.

In the next section we present the model already investigated in Refs. \cite{bazinlo,bazbem}, briefly reviewing its static kink and antikink solutions. We show how this model is related to the $\lambda\phi^4$ model after field and coordinate transformations. We also present the linear stability analysis of the solutions. In the Sec. III we present our main scattering results and then conclude the work in the Sect. IV.


\section{The model}

\begin{figure}
\includegraphics[{angle=0,width=5cm, height=3cm}]{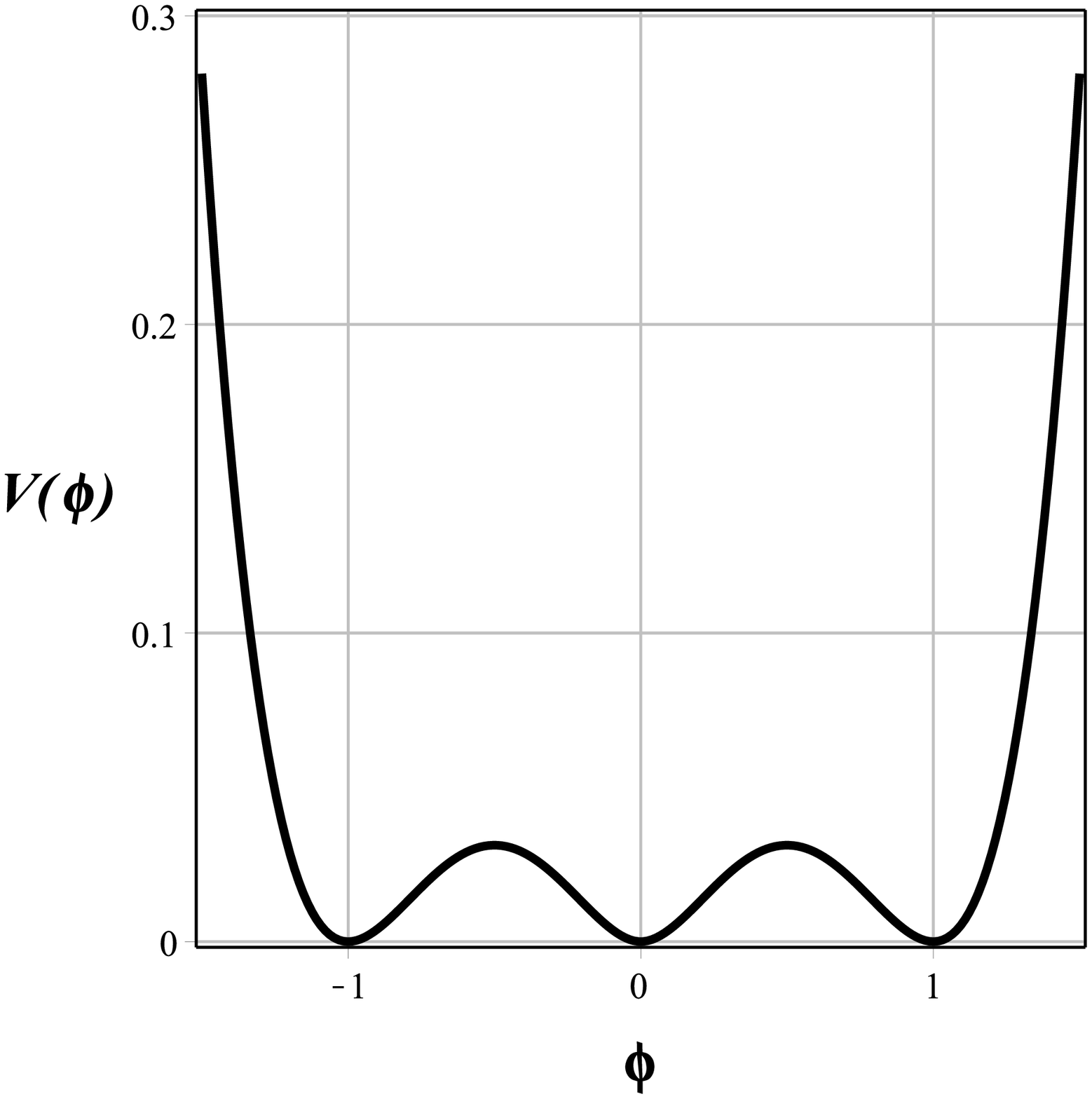}
\includegraphics[{angle=0,width=5cm, height=3cm}]{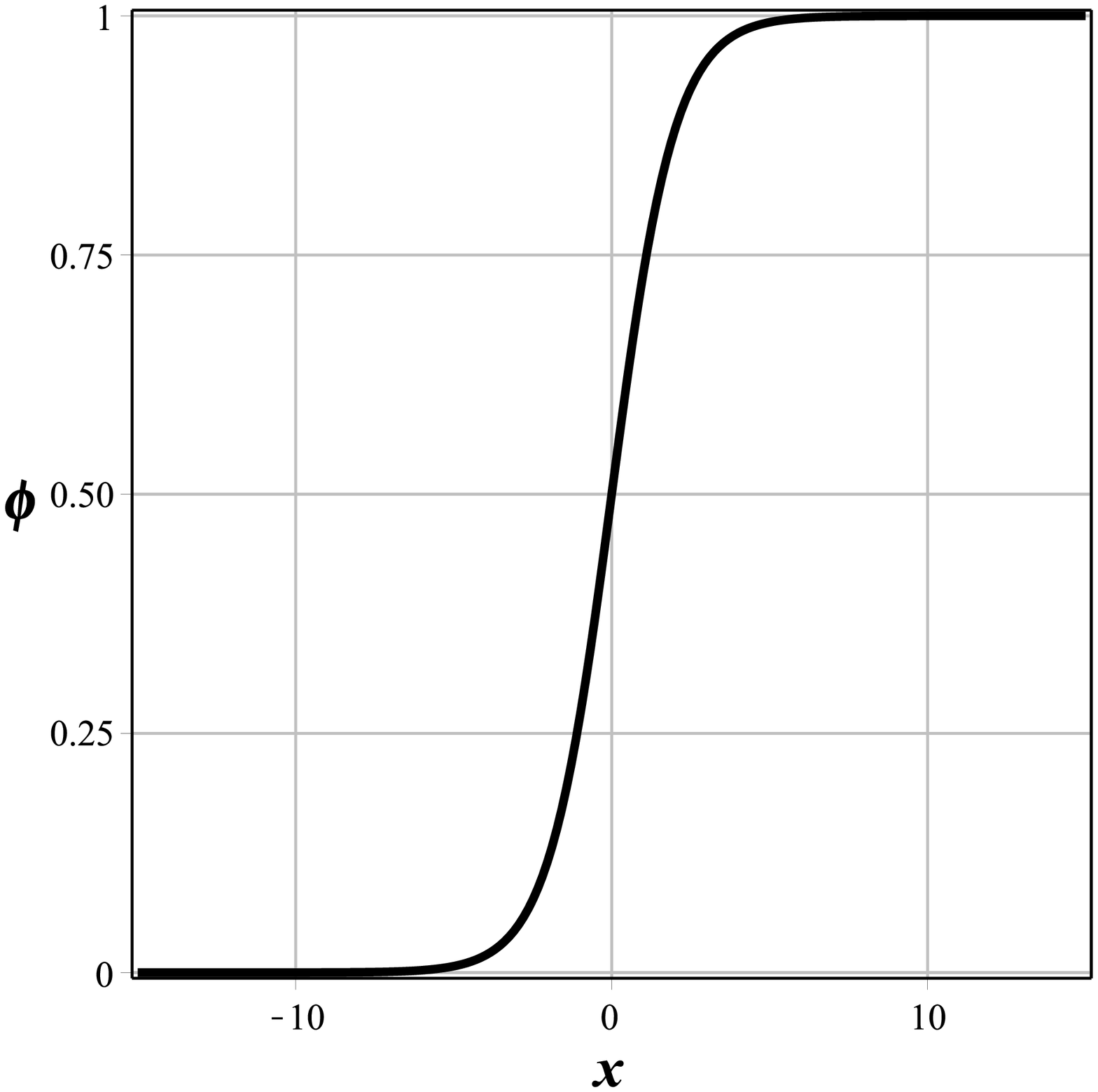}
\includegraphics[{angle=0,width=5cm, height=3cm}]{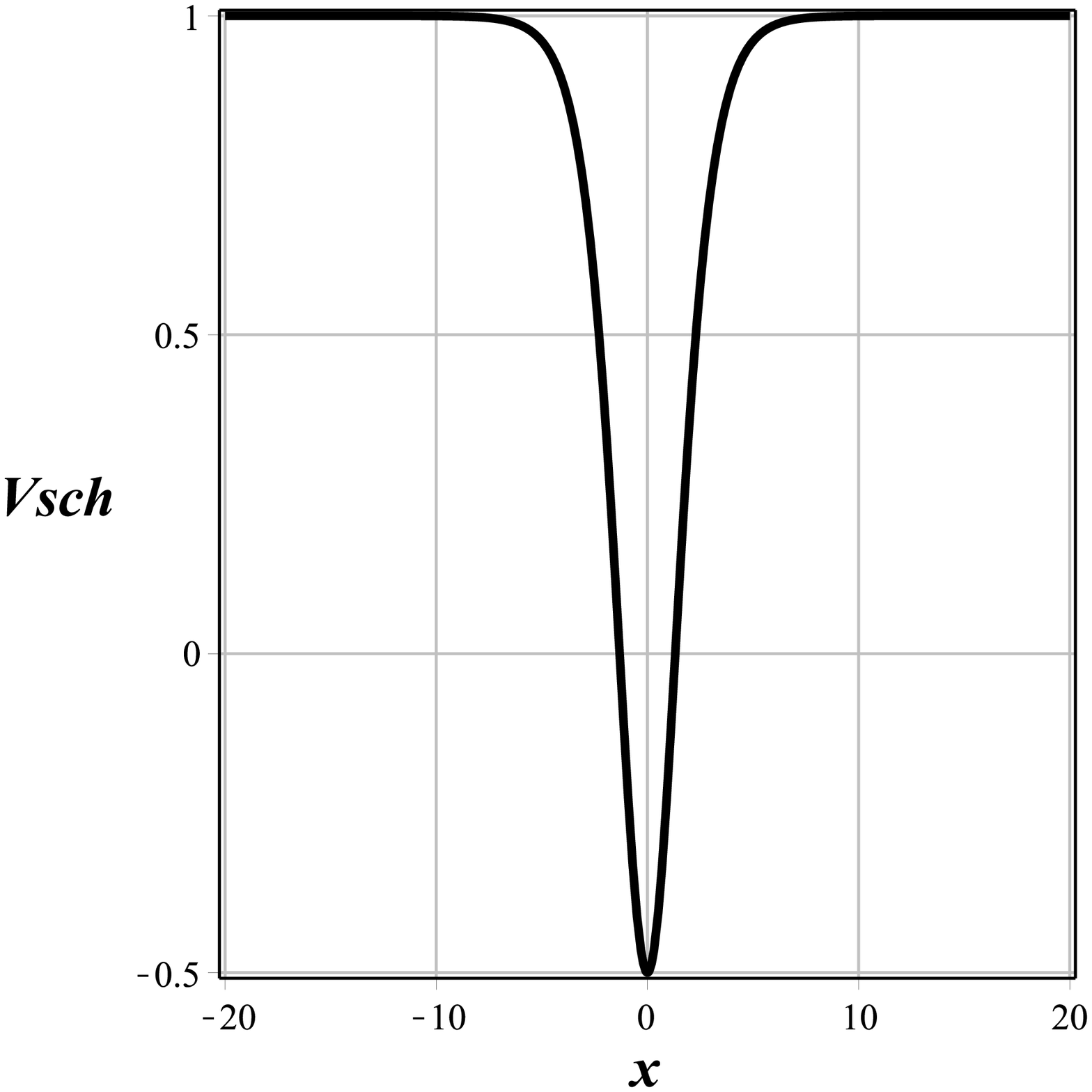}
\caption{(a) Potential $V(\phi)$, (b) Field configuration for the kink $\phi^{(0,1)}$ and c) Schr\"odinger-like potential $V_{sch}(x)$ for perturbations around the kink $\phi^{(0,1)}$. The parameters are fixed as $m^2=\lambda=2$.}
\label{pot_sol}
\end{figure}

Let us consider the action
\begin{equation}
S_{hybrid}=\int  dtdx \biggl( \frac12 \partial_\mu\phi\partial^\mu\phi - V(\phi) \biggr),
\end{equation}
with the potential \cite{bazinlo,bazbem}
\begin{eqnarray}
V(\phi)=\frac{\lambda}{4} \phi^2\biggl(\frac{m}{\sqrt\lambda}-|\phi|\biggr)^2.
\label{potential}
\end{eqnarray}
Fig. \ref{pot_sol}a shows that the potential has three minima in $\phi=0$ and $\phi=\pm \phi_v$, with $\phi_v= {m}/{\sqrt\lambda}$. Then we have two topological sectors connecting adjacent minima. There is symmetry around the minimum $\phi=0$, so the system engenders symmetric $\phi=0$ and asymmetric $\phi=\pm \phi_v$ minima, and can be used to simulate a first-order phase transition, in a way similar to the case of the standard $\phi^6$ model. In the current case, however, given a topological sector, the potential is also symmetric around each one of its local maxima, and this induces an important difference which will make the stability Schr\"odinger-like potential (see below) symmetric, and so well distinct from the stability potential that appears in the $\phi^6$ model, which is asymmetric. The equation of motion is $\ddot\phi-\phi''+dV/d\phi=0$. Static kink solutions are given by
\begin{eqnarray}
  \phi_K^{(0,\phi_v)}(x)&=& \frac12\frac{m}{\sqrt\lambda}\bigg(1+\tanh \bigg(\frac{m}{\sqrt{2}}\frac{x}{2}\bigg) \bigg),\\
  \phi_K^{(-\phi_v,0)}(x)&=& \frac12\frac{m}{\sqrt\lambda}\bigg(-1+\tanh \bigg(\frac{m}{\sqrt{2}}\frac{x}{2}\bigg) \bigg).
   \label{solution}
\end{eqnarray}
Fig. \ref{pot_sol}b depicts the kink profile $\phi_K^{(0,1)}(x)$. Corresponding antikink solutions are given by $\phi_{\bar K}^{(\phi_v,0)}(x)=\phi_K^{(0,\phi_v)}(-x)$ and $\phi_{\bar K}^{(0,-\phi_v)}(x)=\phi_K^{(-\phi_v,0)}(-x)$. Perturbing linearly the scalar field around one kink solution $\phi_K(x)$ as $\phi(x,t)=\phi_K(x)+\eta(x)\cos(\omega t)$ leads to the Schr\"odinger-like equation $-\eta''+V_{sch}\,\eta=\omega^2\eta,$ where the Schr\"odinger-like potential for kinks and antikinks is given by
\be
V_{sch}=\frac{d^2 V}{d\phi^2}=\frac{m^2}2\biggl[-\frac12+\frac32\tanh^2\biggl(\frac{m}{\sqrt{2}}\frac{x}2\biggr)\biggr].
\ee
This stability potential is presented in Fig. \ref{pot_sol}c. 

Note that the hybrid model has deep connections to the well-studied $\lambda\phi^4$ model. Indeed, for $\phi>0$  and after the transformations $\phi\to\frac12({m}/{\sqrt\lambda}+\phi)$, $ x^\mu\to 2x^\mu$,
 we get $S_{hybrid}=\frac14S_{\lambda\phi^4}$, where $S_{\lambda\phi^4}$ is the action for the $\lambda\phi^4$ model with potential $
V(\phi)=\frac\lambda4({m^2}/{\lambda}-\phi^2)^2$ and corresponding kink solution $\phi=\frac{m}{\sqrt\lambda}\tanh( {m}x/{\sqrt2} ). $
Analogously, the same model and solution is recovered for $\phi<0$ under the transformations
$\phi\to\frac12(-{m}/{\sqrt\lambda}+\phi)$, $ x^\mu\to 2x^\mu$. The equation for perturbations for the hybrid model can be mapped, with $x^\mu\to 2x^\mu$, to the equation of perturbations for the $\lambda\phi^4$ model, where the corresponding Schr\"odinger-like potential is the Poschl-Teller with known \cite{sug} eigenvalues and eigenfunctions. The eigenvalues from both models are related as $\omega^2_{hibrid}=\frac14\omega^2_{\lambda\phi^4}$. The Schr\"odinger-like equation for the hybrid model has two bound eigenstates: the zero-mode or translational state, a vibrational mode: 
\begin{eqnarray}
\omega^2_0 &=& 0, \,\,\,\,\, \eta_0=\sqrt{\frac38\frac{m}{\sqrt2}}\sech^2\biggl(\frac{m}{\sqrt2}\frac x2\biggr), \\
\label{omega1}
\omega^2_1 &=& \frac38m^2, \,\,\,\,\, \eta_1=\sqrt{\frac34\frac{m}{\sqrt2}}\tanh\biggl(\frac{m}{\sqrt2}\frac x2\biggr) \sech\biggl(\frac{m}{\sqrt2}\frac x2\biggr). 
\end{eqnarray}  
There is also a continuum of states described by
\begin{eqnarray}
\label{omegak}
\omega^2_k &=& \frac14({k^2}+2m^2), \\
\eta_k&=&N_ke^{ik\frac x2}
\biggl[ 3 \tanh^2\biggl(\frac{m}{\sqrt2}\frac x2\biggr) -1 - \frac2{m^2}k^2 - 3\sqrt2i \frac{k}m\tanh\biggl(\frac{m}{\sqrt2}\frac x2\biggr) \biggr], 
\end{eqnarray}
with 
\be
N_k^{-2}=8\pi \biggl[ 2 \biggl( \frac{k^2}{m^2} +1 \biggr)^2 + \frac{k^2}{m^2}  \biggr].
\ee
The continuum states are normalized as \cite{sug}
\be
\int dx \eta_{k'}^*(x) \eta_k(x) = \delta(k'-k). 
\ee

Note that, despite strictly connected, the hybrid and $\lambda\phi^4$ models are not the same, since there is no unique transformation that can be applied to bring one to another. However, even being different, many properties of the hybrid model are inherited from the $\lambda\phi^4$ model. One can cite the simple rescaling between the energy eigenvalues, the simple connection between the eigenfuntions and a $Z_2$-symmetric Schr\"odinger-like potential. As occurs in the $\lambda\phi^4$ model, the Schr\"odinger-like potential for the hybrid model is the same for kink-antikink and antikink-kink solutions.  This implies, in the $\lambda\phi^4$ model, that the scattering process is the same for both configurations. We will see that in the hybrid model, on the contrary, kink-antikink and antikink-kink have different structures. This is connected to the possibility of changing the topological sector at the outcome of the collision process. The numerical analysis of Schr\"odinger-like potentials for kink-antikink and antikink-kink  show that negative eigenvalues are absent, meaning that the configurations are stable. This is corroborated by the scattering analysis presented in the following section, since each initial pair configuration travels without loosing energy before the interaction. 

From here on, and without loosing generality, we will consider the initial profile belonging to the topological sector connecting the vacua $\phi=0$ and $\phi=\phi_v$. This is justified because the Schr\"odinger-like potential is symmetric.  


\section { Numerical Results }


Here we present our main results of antikink-kink and kink-antikink scattering. We solved the equation of motion with a  pseudospectral method on a grid with $2048$ nodes with periodic boundary conditions. We fixed $x=\pm x_0$ with $x_0 = 15$ for the initial symmetric position of the pair and set the grid boundary at $x=\pm x_{max}$ with $x_{max}=200$. A sympletic method with the Dirichlet condition imposed at the boundaries was also applied to double check our numerical results. We used a $4^{th}$ order finite-difference method with spatial step $\delta x=0.09$ and a $6^{th}$ order symplectic integrator with time step $\delta t=0.04$. In this section we fix the parameters $\lambda=m^2=2$.


\subsection{Antikink-kink ($\bar K K$) collisions}


\begin{figure}
\includegraphics[{angle=0,width=4cm,height=4cm}]{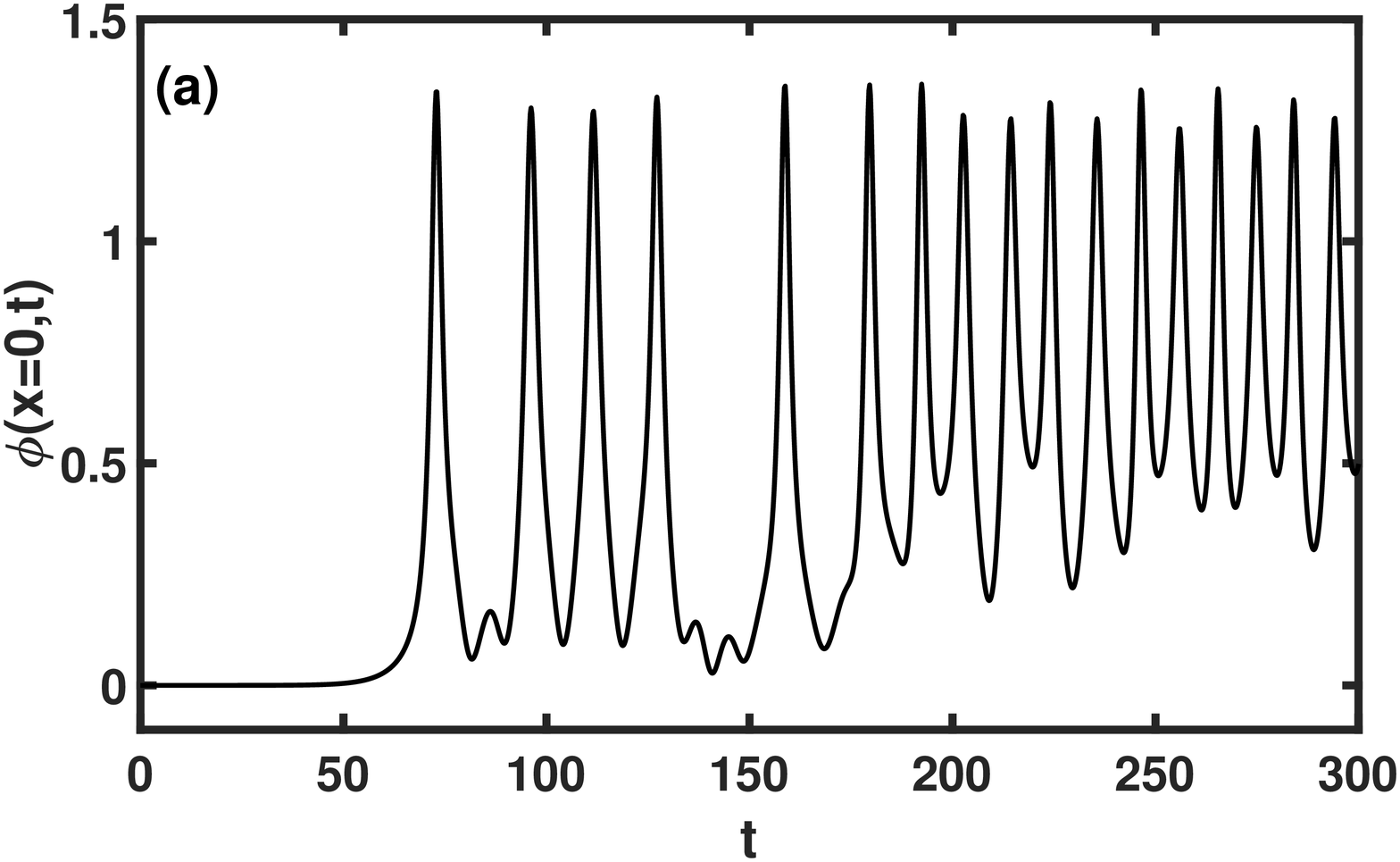}
\includegraphics[{angle=0,width=4cm,height=4cm}]{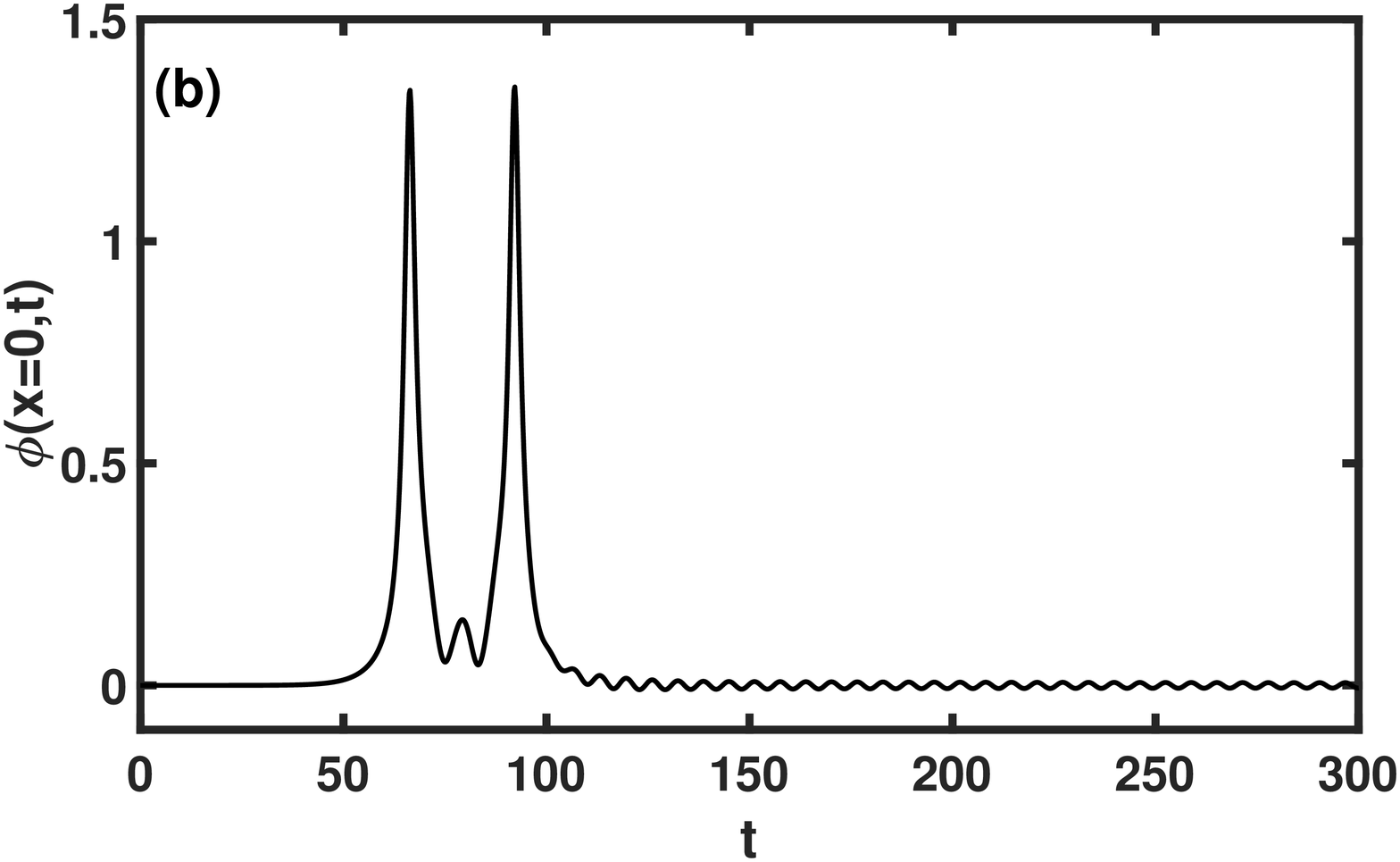}
\includegraphics[{angle=0,width=4cm,height=4cm}]{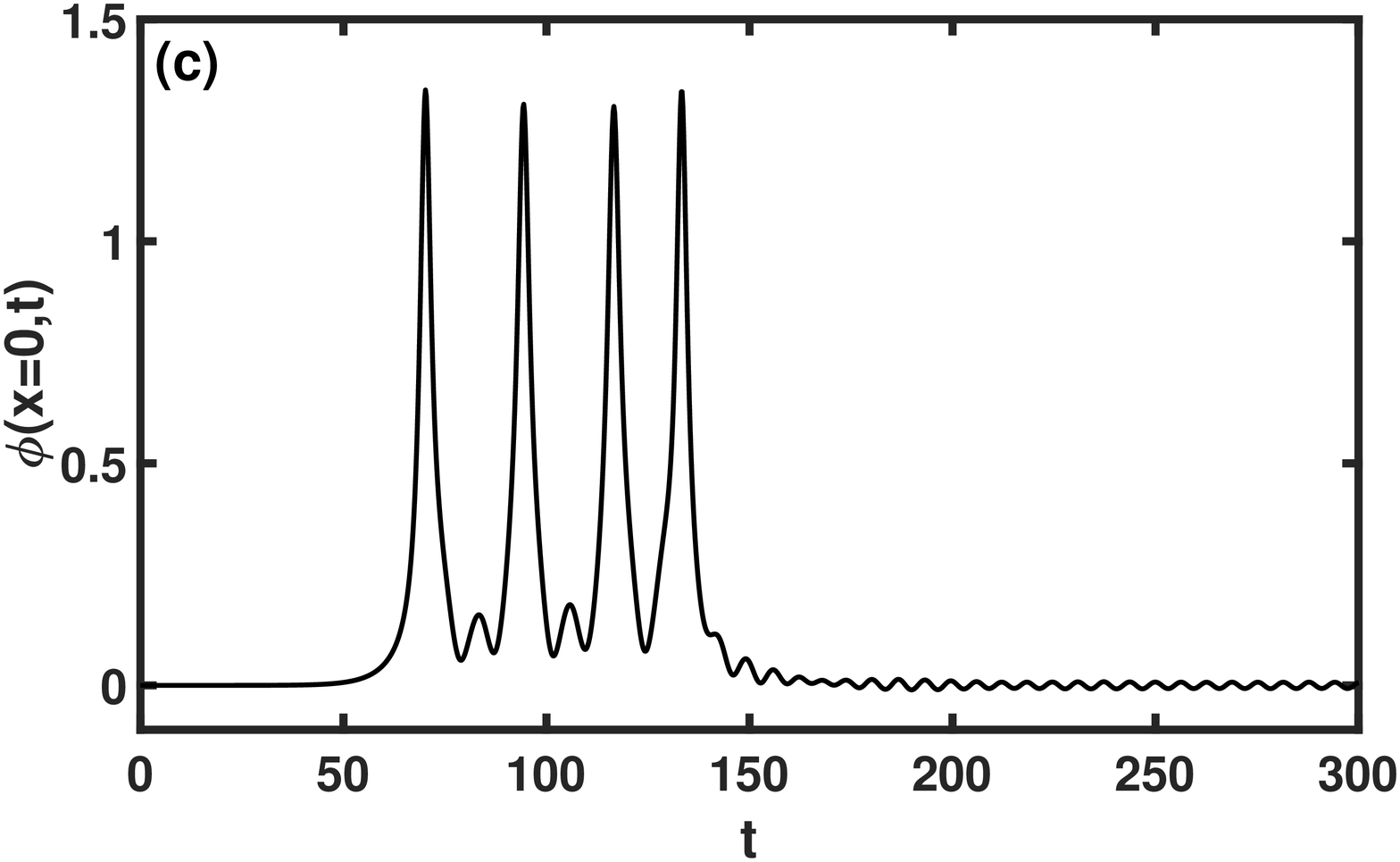}
\includegraphics[{angle=0,width=4cm,height=4cm}]{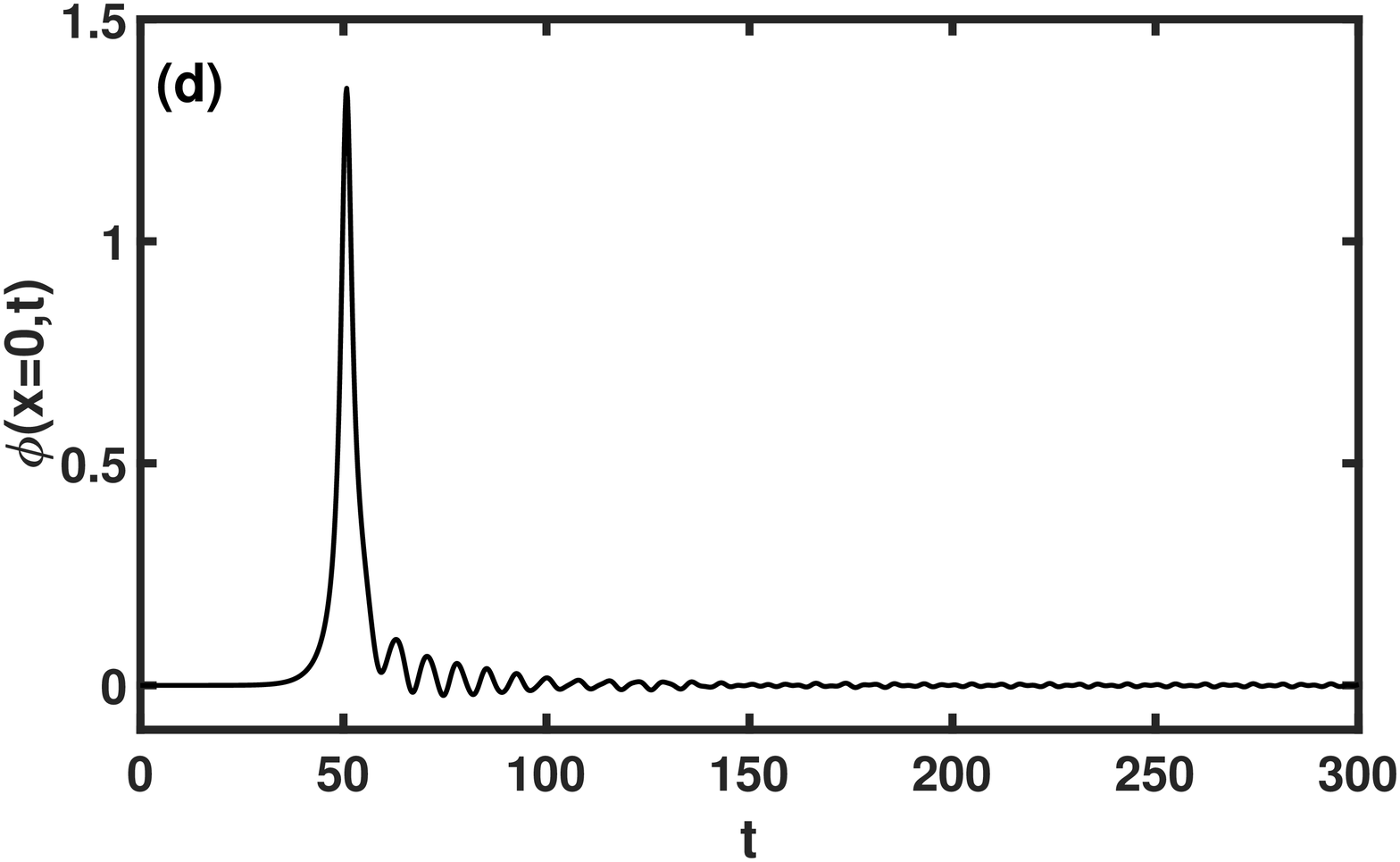}
\\
\includegraphics[{angle=0,width=4cm,height=4cm}]{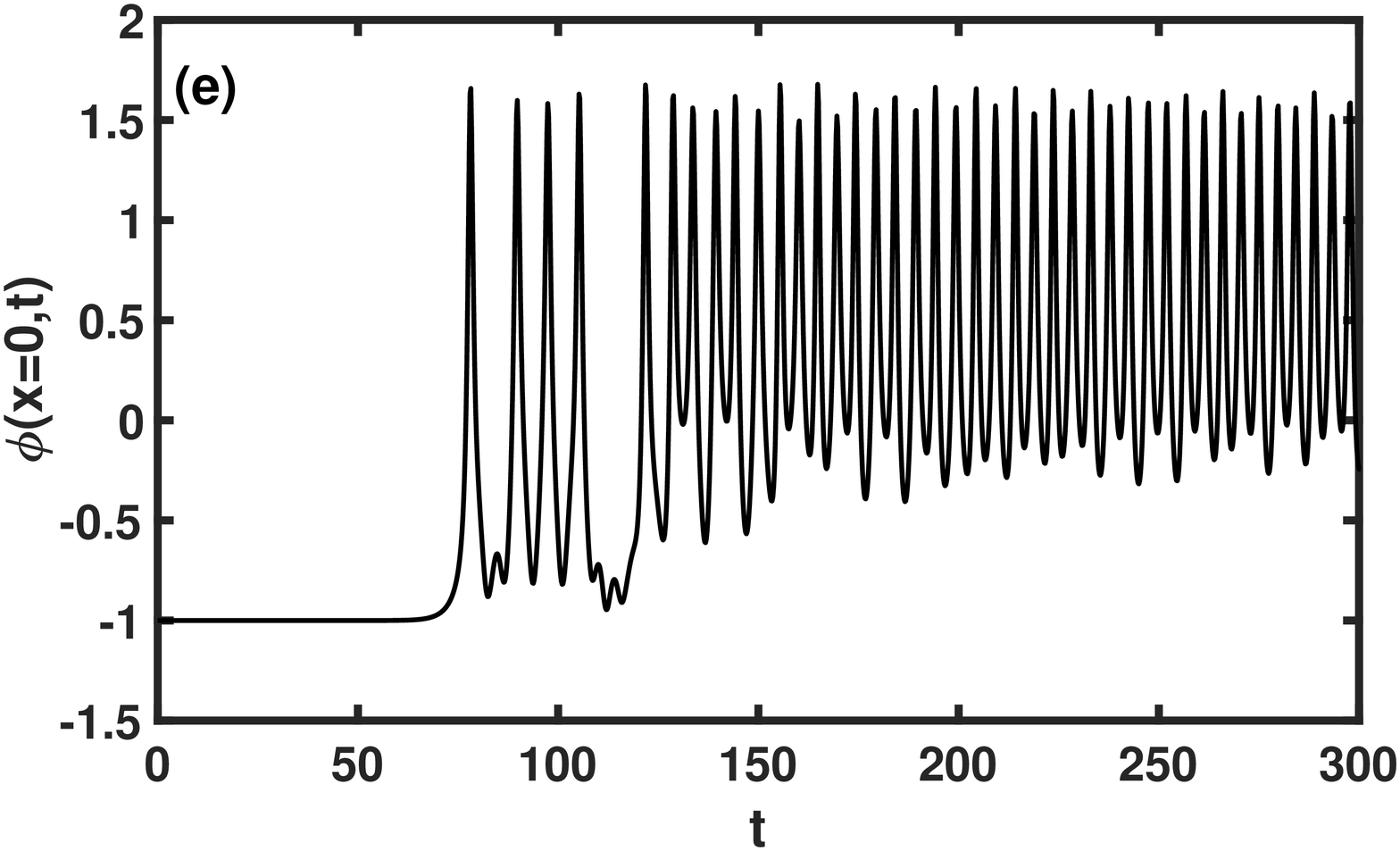}
\includegraphics[{angle=0,width=4cm,height=4cm}]{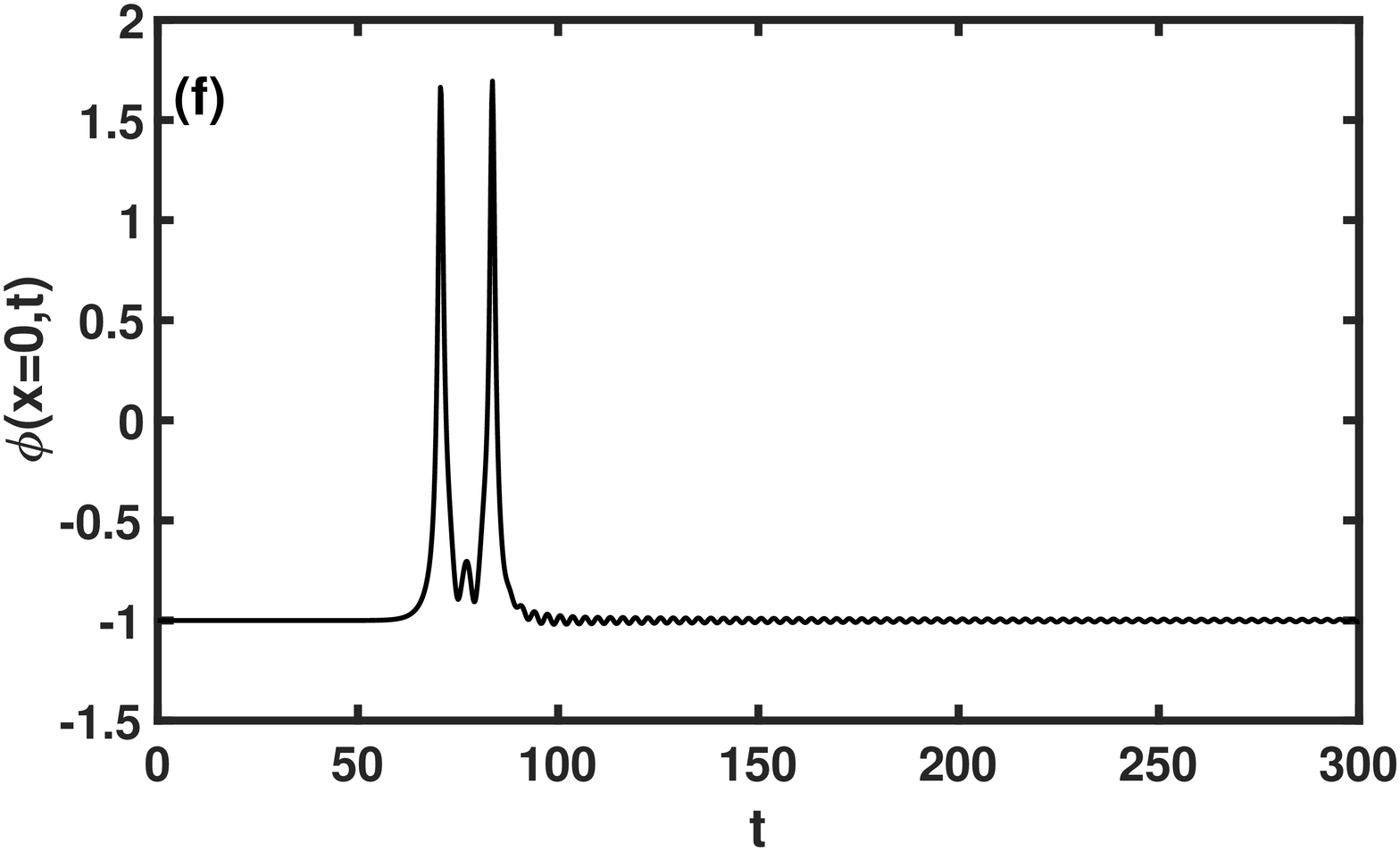}
\includegraphics[{angle=0,width=4cm,height=4cm}]{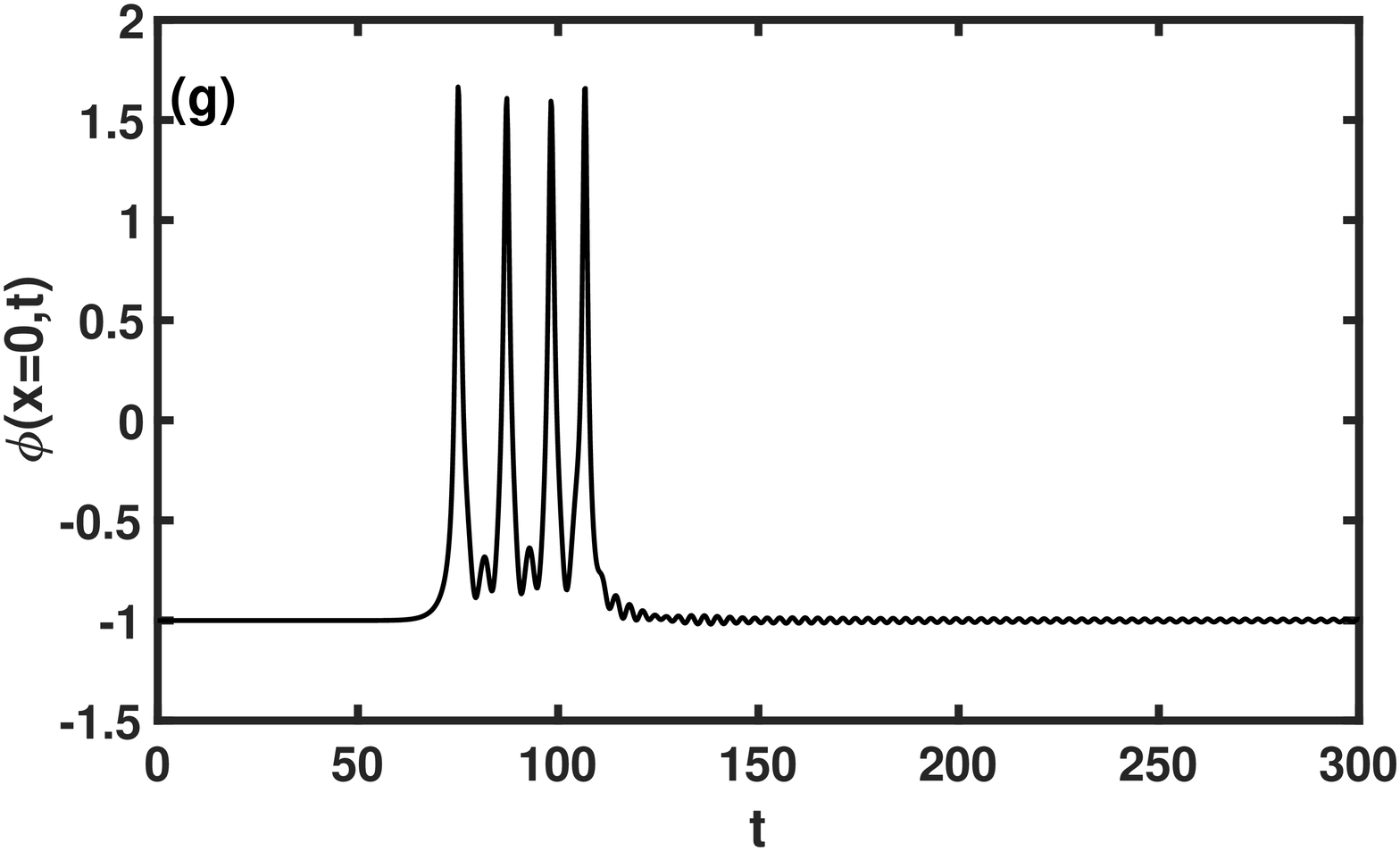} 
\includegraphics[{angle=0,width=4cm,height=4cm}]{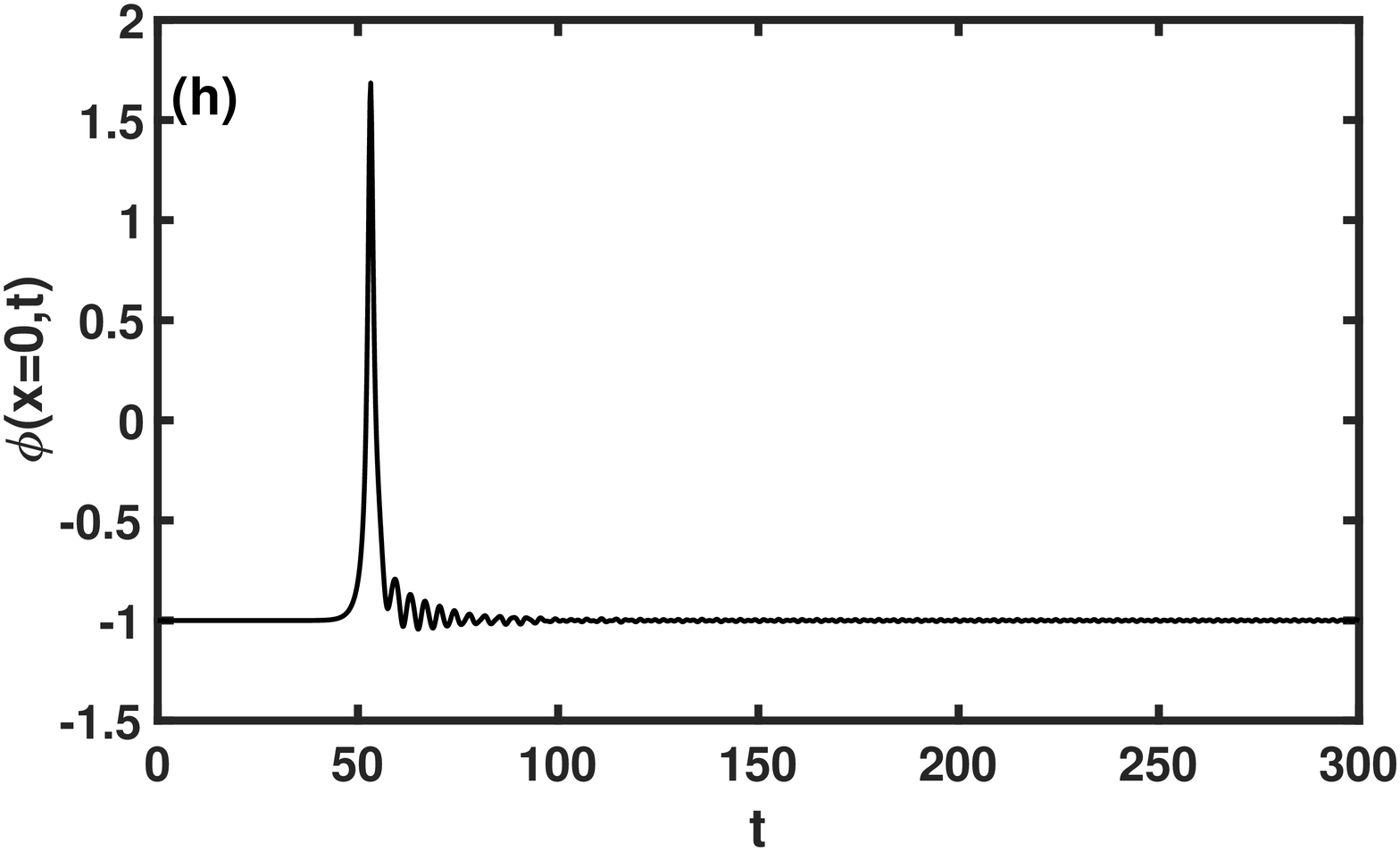}
\caption{Antikink-kink collisions: $\phi(x = 0, t)$ for the hybrid model (higher figures) and for the $\phi^4$ model (lower figures), showing (a), (e) bion state for $v=0.18$, (b), (f) two-bounces for $v=0.2$, (c), (g) four-bounces for $v=0.1877$ and (d), (h) one-bounce for $v=0.27$. }
\label{AK}
\end{figure}

To solve the equation of motion for antikink-kink scattering we use the following initial conditions: $\phi(x,0)=\phi_{\bar{K}}(x+x_0,v,0)+\phi_K(x-x_0,-v,0)$ and $\dot\phi(x,0)=\dot\phi_{\bar{K}}(x+x_0,v,0)+\dot\phi_K(x-x_0,-v,0)$ where $\phi_{\bar{K}}(x+x_0,v,t)$ means a Lorentz boost solution for antikink with velocity $v$, centered at $x=-x_0$.
 For $v<v_c$ with $v_c = 0.2599$, bion states are achieved, where the scalar field at the center of mass $\phi(0,t)$ changes after the scattering from the initial value $\phi=0$ to erratic oscillations around the adjacent vacuum $\phi=1$, as in the example shown in the Fig. \ref{AK}a. After long time emiting scalar radiation, the antikink-kink pair annihilates and the scalar field goes to
 the vacuum $\phi=1$. For $v>v_c$ the output is an inelastic scattering between the pair. In this case, $\phi(0,t)$ shows
 one-bounce (represented by $N_b=1$) between the vacuum $\phi=0$ - see, for instance, the Fig. \ref{AK}d. Also, for some windows in velocities $v \lesssim v_c$, $\phi(0,t)$ presents two-bounce ($N_b=2$) between the vacuum $\phi=0$, as in the example shown in the Fig. \ref{AK}b. Close to two-bounce windows there appear three-bounce windows. This process repeats in a fractal way with higher-bounce windows. One example of a collision with four-bounce is depicted in the Fig. \ref{AK}c. For comparison we included in the  Figs. \ref{AK}e-h,  the results for the $\phi^4$ model with same initial velocities used in 
Figs. \ref{AK}a-d for the hybrid model.

Note from the examples of the figures Fig. \ref{AK}a (bion), \ref{AK}b (two-bounces) and \ref{AK}c (four-bounces) that the scalar field do not cross to the other topological sector during and after the collision. Then the mapping between hybrid and $\phi^4$ models is justified and the phenomenological CSW mechanism can be used to understand the presence of two-bounce windows as a resonant mechanism described by $ \omega_1T'=2\pi m+\theta_1$, where $T'$ is the time interval between the bounces and $\theta_1$ is a phase shift.  This means that, for collisions belonging to the same windows, and since  $\omega_1^{(hybrid)}=\omega_1^{(\phi^4)}/2$, the time interval between the bounces for the hybrid model is twice larger, in comparison to the $\phi^4$ model. This can be verified comparing, for instance, Figs. \ref{AK}b and \ref{AK}f. The figure \ref{AK}d for the one-bounce shows oscillations of the scalar field for negative values of $\phi$ after the scattering. This can be interpreted as the scalar radiation emitted by the antikink-kink pair which can be described in terms of the frequencies of continuum mode - smaller for the hybrid model as $\omega_k^{(hybrid)}=\omega_k^{(\phi^4)}/2$. 
The same reduction of the frequency of oscillations, described above  for two-bounce and one-bounce collisions, is also observed for the bion states (compare  Figs. \ref{AK}a and \ref{AK}e). The unifying reason for this is in the transformation of coordinates $x^{\mu} \to 2 x^{\mu}$ that connect both models.

\begin{figure}
\includegraphics[{angle=0,width=8cm, height=5cm}]{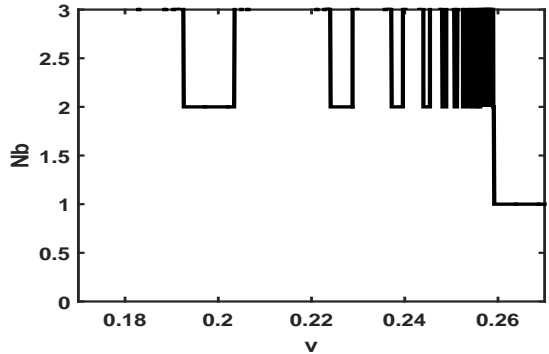}
\caption{Antikink-kink collisions: number of bounces $N_b$ versus initial velocity $v$ showing expected two-bounce windows according to CSW mechanism.}
\label{AK-windows}
\end{figure}

Considering that in between the two-bounces the scalar field oscillates near to the initial vacuum $\phi=0$, we have investigated if a crossing to the other topological sector could appear for collisions with higher number of bounces. We found no signal of crossing for collisions with three- and four-bounces. This signals that for antikink-kink collisions the hybrid and $\phi^4$ models can indeed be mapped. Naturally, despite mapped, the models are different, with detectable differences on the scattering, as we showed in the Fig. \ref{AK}. One further aspect to be explored here is the structure of bounce-windows. Fig. \ref{AK-windows} summarizes our main results for the number of bounces as a function of initial velocity. Note that the thickness of each two-bounce windows decreases with the velocity as one approaches the limit $v=v_c$ from bellow. The order of the 2-bounce windows is the number $m$ of oscillations between the bounces. For instance, Fig. \ref{AK}b shows a plot of $\phi(0,t)$ with $m=1$, belonging to the first two-bounce window. Comparing the scattering results for the hybrid model with those for the $\lambda\phi^4$ model \cite{aom}, we see that the Fig. \ref{AK-windows} roughly matches the Fig. 3a from Ref. \cite{twin} for the $\lambda\phi^4$ model. We also studied the structure of some three-bounce windows. We observed that the extrema that define the interval in velocity are not the same. Despite of this, their length are equal, considering the round error.


\subsection{Kink-antikink ($K\bar K$) collisions}


In this case the initial conditions are given by $\phi(x,0)=\phi_K(x+x_0,v,0)+\phi_{\bar{K}}(x-x_0,-v,0)-{m}/{\sqrt{\lambda}}$ and $\dot\phi(x,0)=\dot\phi_K(x+x_0,v,0)+\dot\phi_{\bar{K}}(x-x_0,-v,0).$
\begin{figure}
\includegraphics[{angle=0,width=6cm}]{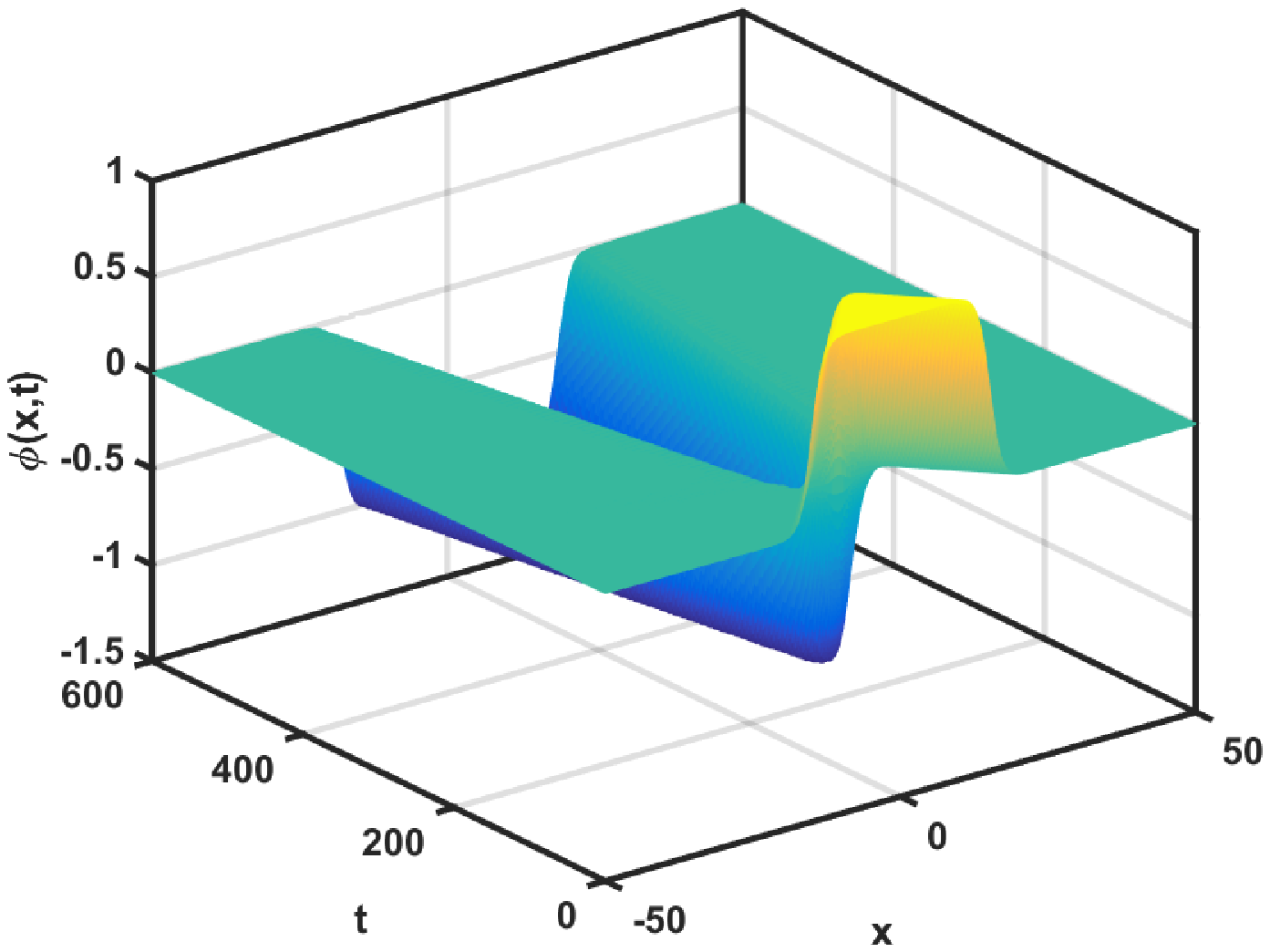} 
\includegraphics[{angle=0,width=6cm}]{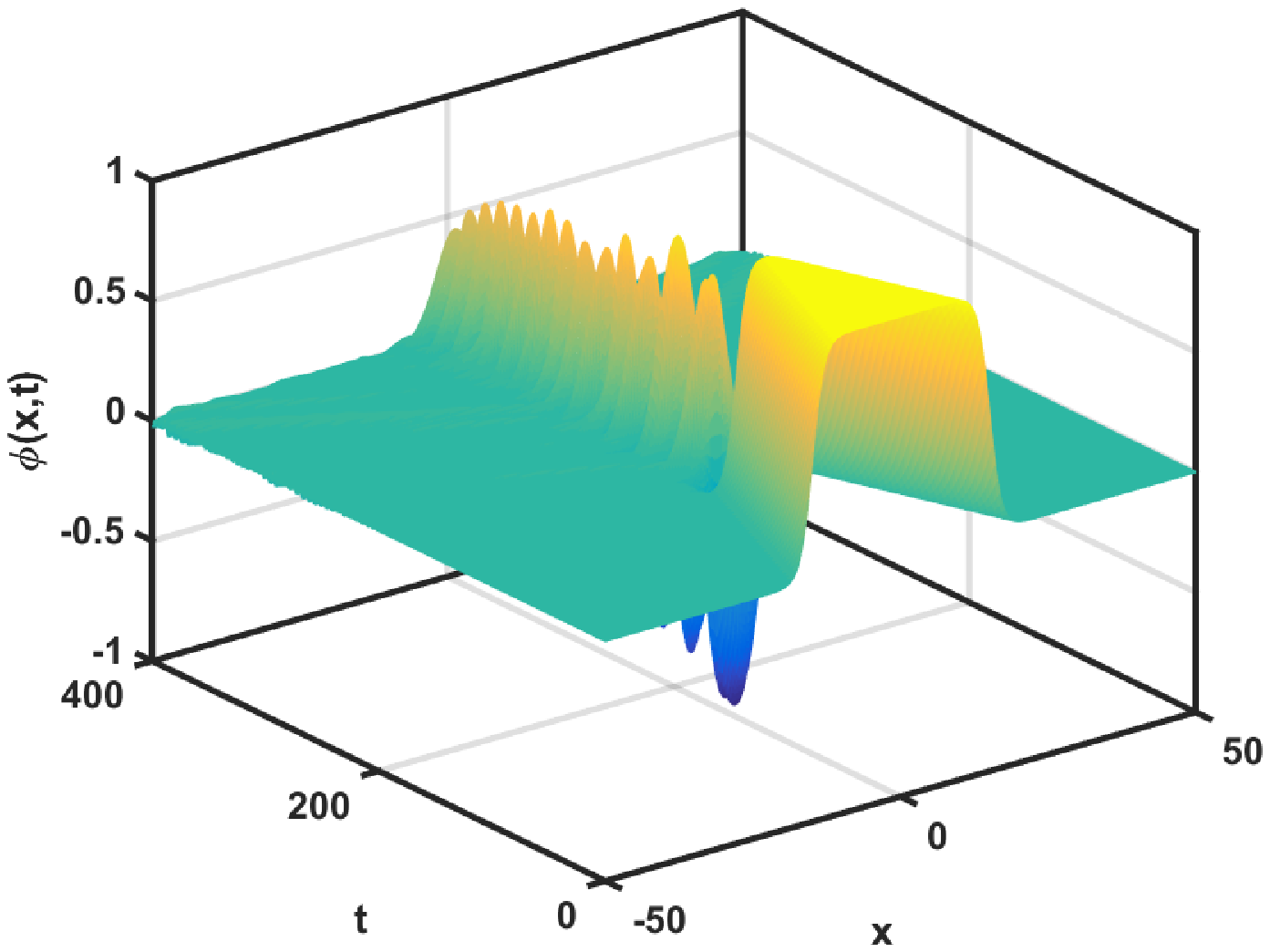} 
\caption{(a) Scalar field $\phi(x,t)$ for the kink-antikink collision at high velocity $v>v_{crit}$  (here $v=0.154$). Note the change of phase after the solution. (b) Bion state for $v<v_{crit}$  (here $v=0.09$).}
\label{bounce_2}
\end{figure}
We analyzed the collisions varying the initial velocity $v$. For $v>v_{crit}$, with $v_{crit}=0.152$, the scalar field gains a phase shift changing the topological sector as $\phi_K^{(0,1)}(x,t)+\phi_{\bar K}^{(1,0)}(x,t) \rightarrow \phi_{\bar K}^{(0,-1)}(x,t)+\phi_K^{(-1,0)}(x,t)$, as depicted in Fig. \ref{bounce_2}a. There we see that the scalar field at the center of mass changes abruptly from the vacuum $\phi=1$ to the vacuum $\phi=0$. For most velocities $v<v_{crit}$ we have bion states and the scalar field at the center of mass, initially in the vacuum $\phi=1$, oscillates erratically after the collision around the vacuum $\phi=0$ - see Fig. \ref{bounce_2}b.
\begin{figure}
\includegraphics[{angle=0,width=5cm}]{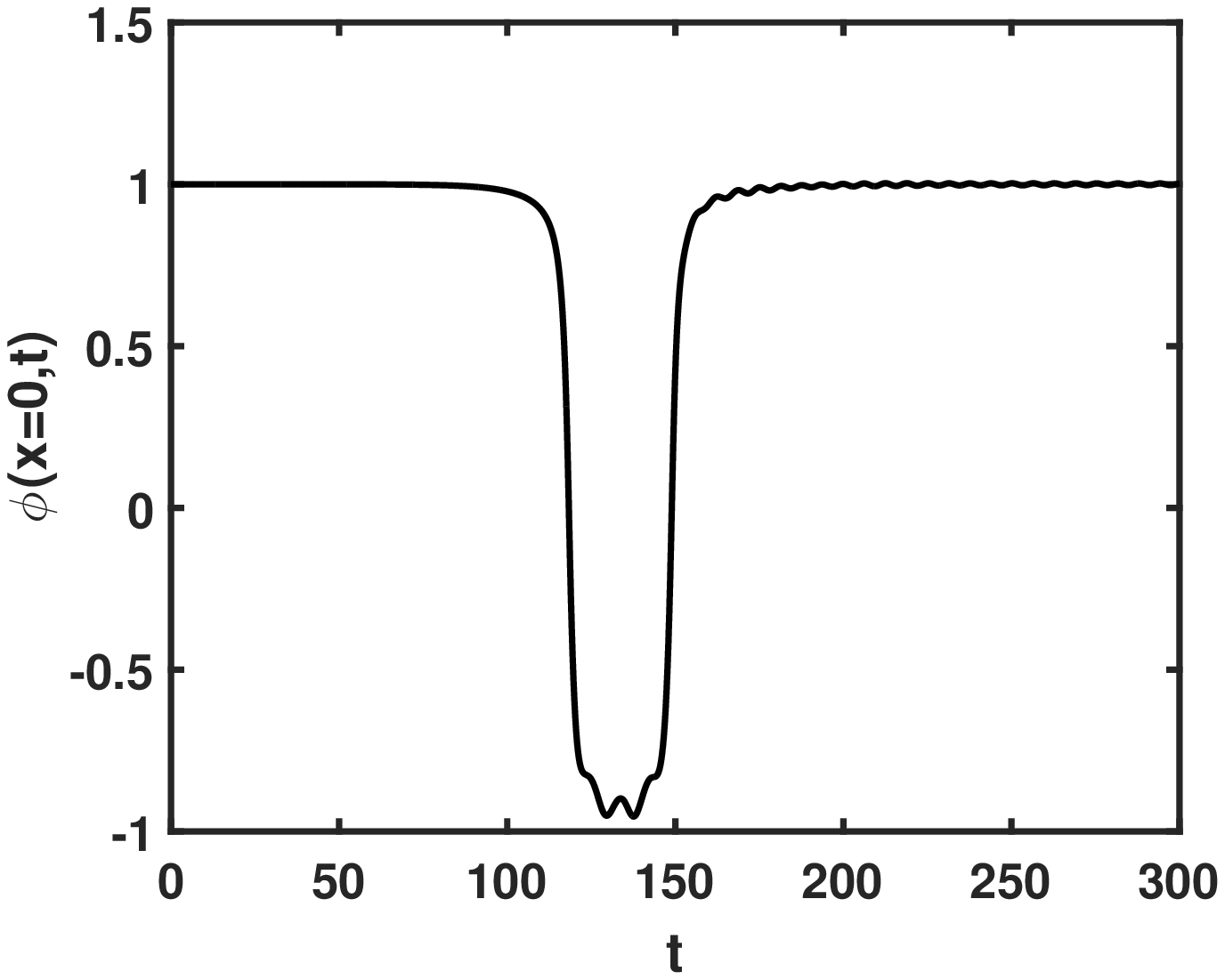}
\includegraphics[{angle=0,width=5cm}]{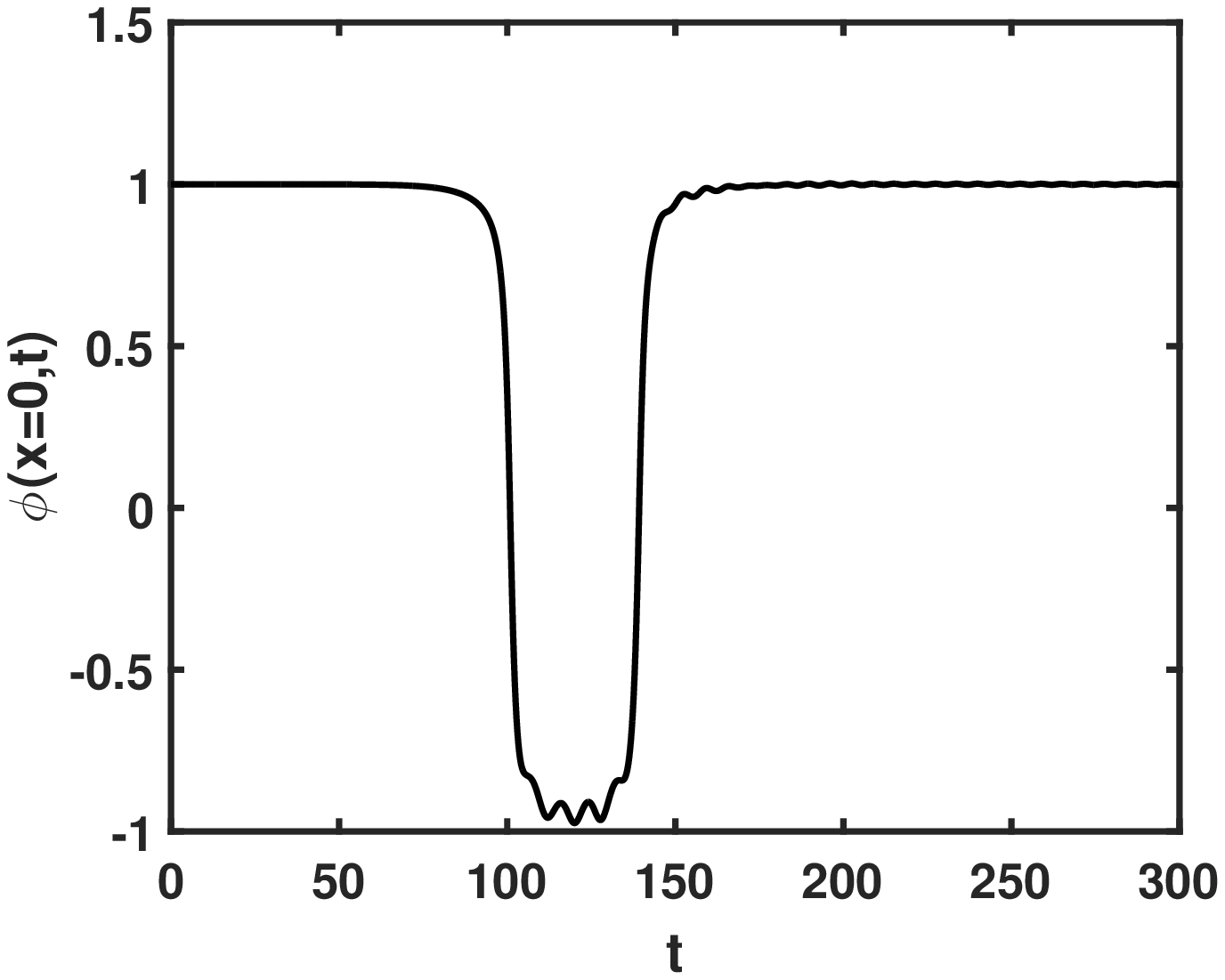} 
\includegraphics[{angle=0,width=5cm}]{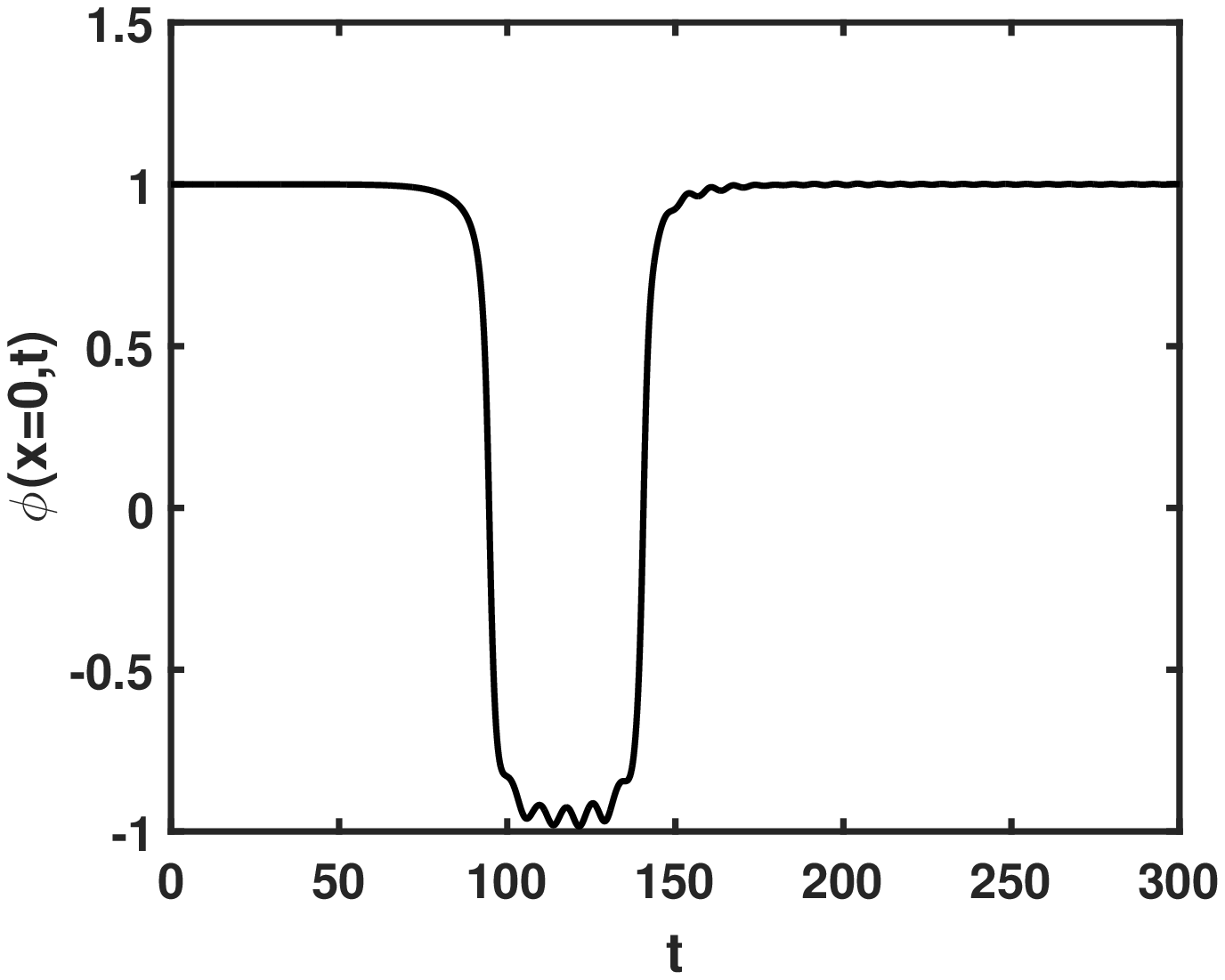}
\caption{Kink-antikink collisions: scalar field at the center of mass $\phi(x = 0, t)$ versus time for (a) $v=0.104$, with $N=2$ oscillations during the one-bounce;  (b) $v=0.124$ with $N=3$;  (c) $v=0.133$ with $N=4$.}
\label{bounce_1}
\end{figure}
For some velocities $v \lesssim v_c$, despite linear perturbations leading to vibrational states for both kink and antikink, there is no evidence of two-bounces like the one described in the Fig. \ref{AK}b for antikink-kink collisions. There we saw that $\phi(0,t)$ oscillates around the initial vacuum $\phi=0$. In the present case, on the contrary,  we have cases in which the scalar field, initially in the vacuum $\phi=1$, bounces once; during the bouncing the scalar field presents a certain number $N$ of oscillations in the other topological sector around $\phi=-1$ - see Figs. \ref{bounce_1}a-c. This pattern was already observed in the modified sine-Gordon model \cite{peycam}. Fig. \ref{KA}(a) shows the distribution of one-bounce windows in a plot of $N$ versus the initial velocity. Note from the figure the presence of one-bounce windows with $N$ growing and their thickness decreasing with $v$. In particular, the Figs. \ref{bounce_1}a-c correspond to the first three one-bounce windows from Fig. \ref{KA}(a).

\begin{figure}
\includegraphics[{angle=0,width=8cm, height=4cm}]{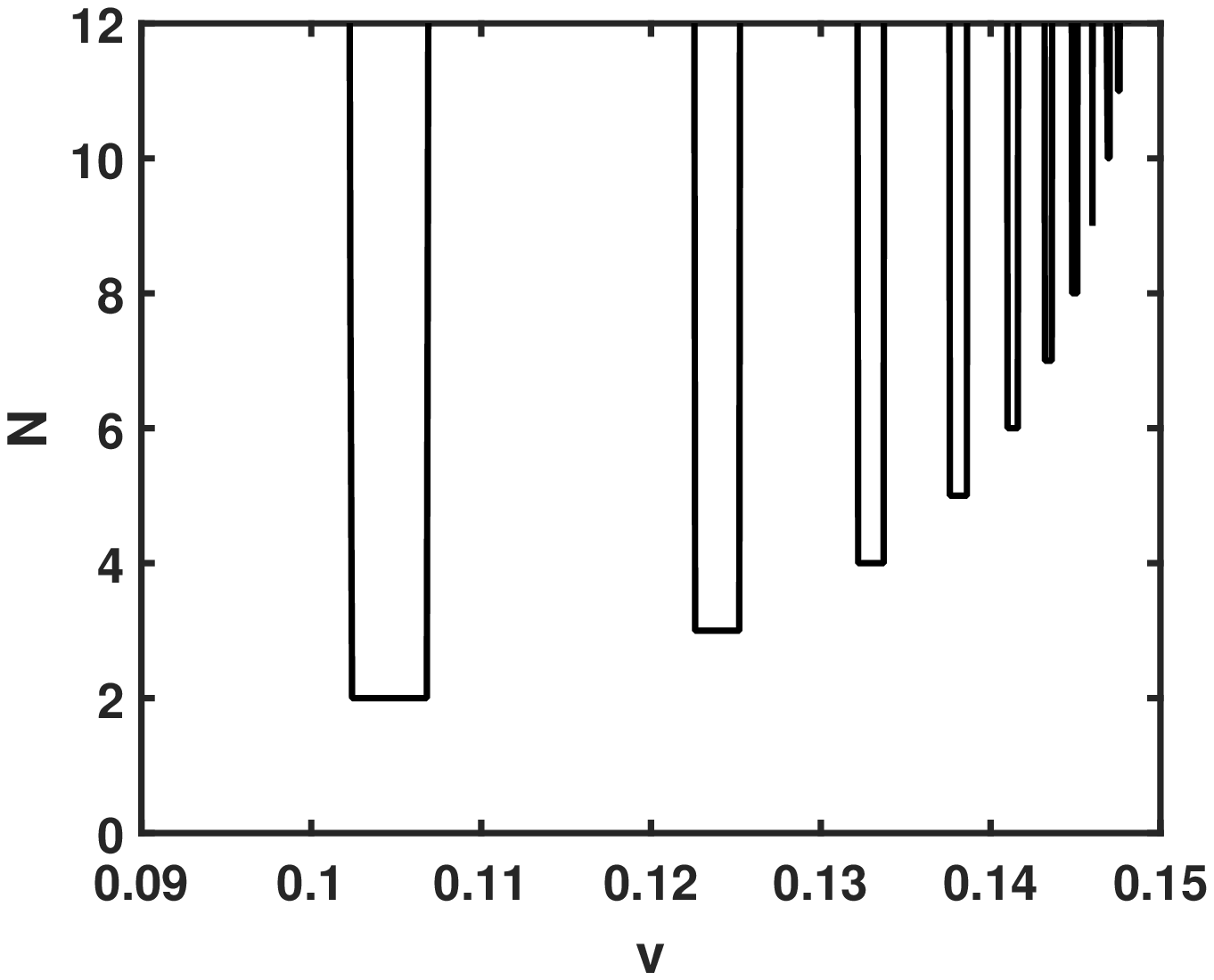}
\includegraphics[{angle=0,width=8cm, height=4cm}]{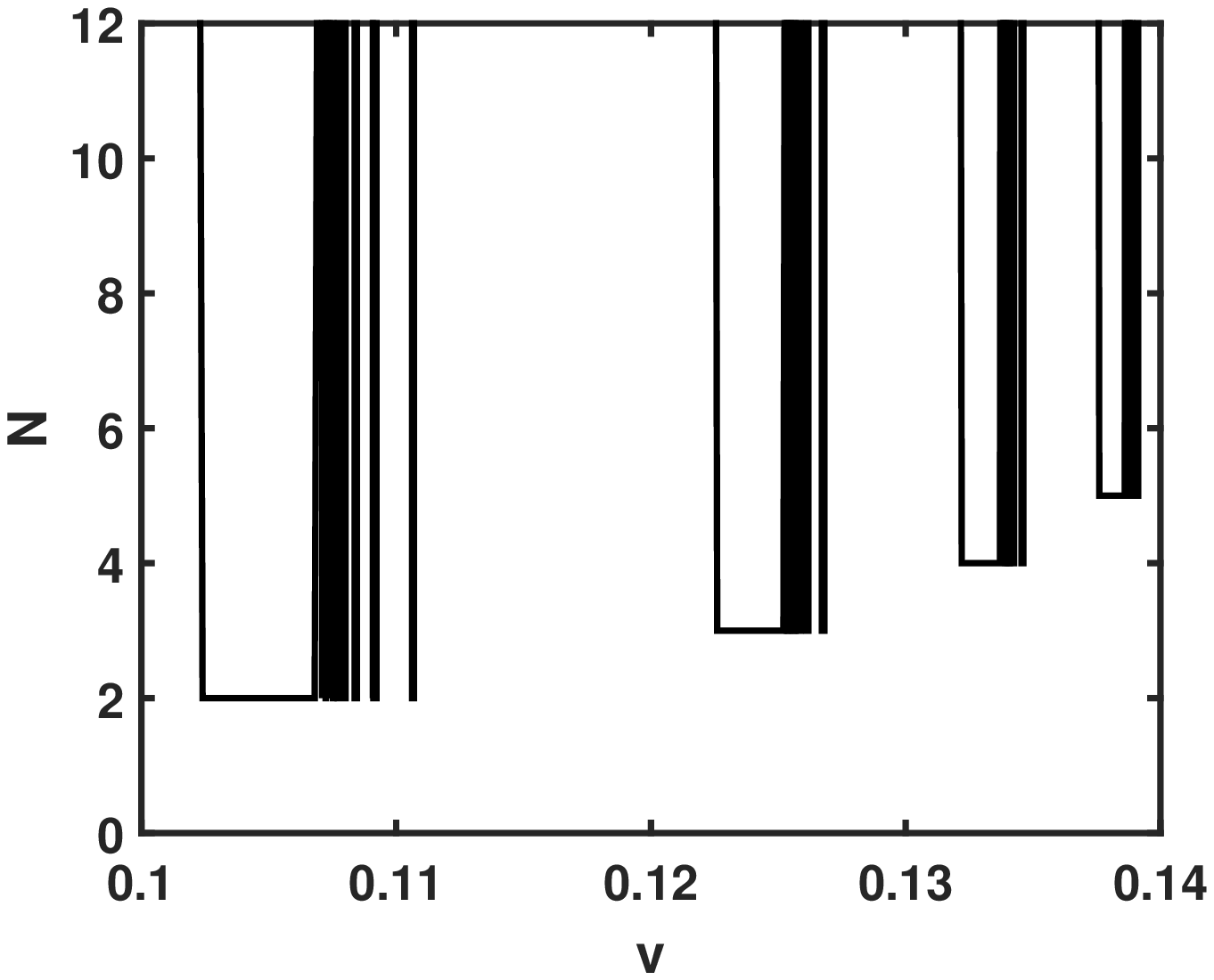}
\caption{Kink-antikink collisions: a) number $N$ of oscillations during an one-bounce collision versus initial velocity. b) Close to the one-bounce windows there is a series of thinnner windows with one-bounce collision followed by a change to the other vacuum state. }
\label{KA}
\end{figure}

The occurrence of oscillations in the one-bounce collisions can be explained
as a mechanism of resonance: initially the pair has its energy in the translational mode; during the oscillations the energy is stored in the vibrational mode. After some oscillations the pair is released following a relation of the form $\omega_1T=2\pi N+\theta_2$, where $T$ is the time interval of the one-bounce and  $\theta_2$ is a phase shift. This is similar to the CSW mechanism described before for two-bounce windows for antikink-kink collisions, trading i) $T'$ by $T$ and ii) $m$, the number of oscillations between the bounces,  by $N$. The measured slope $7.41$ is close to the expected value $2\pi/\omega_1=7.26$.

\begin{figure}
\includegraphics[{angle=0,width=5cm, height=5cm}]{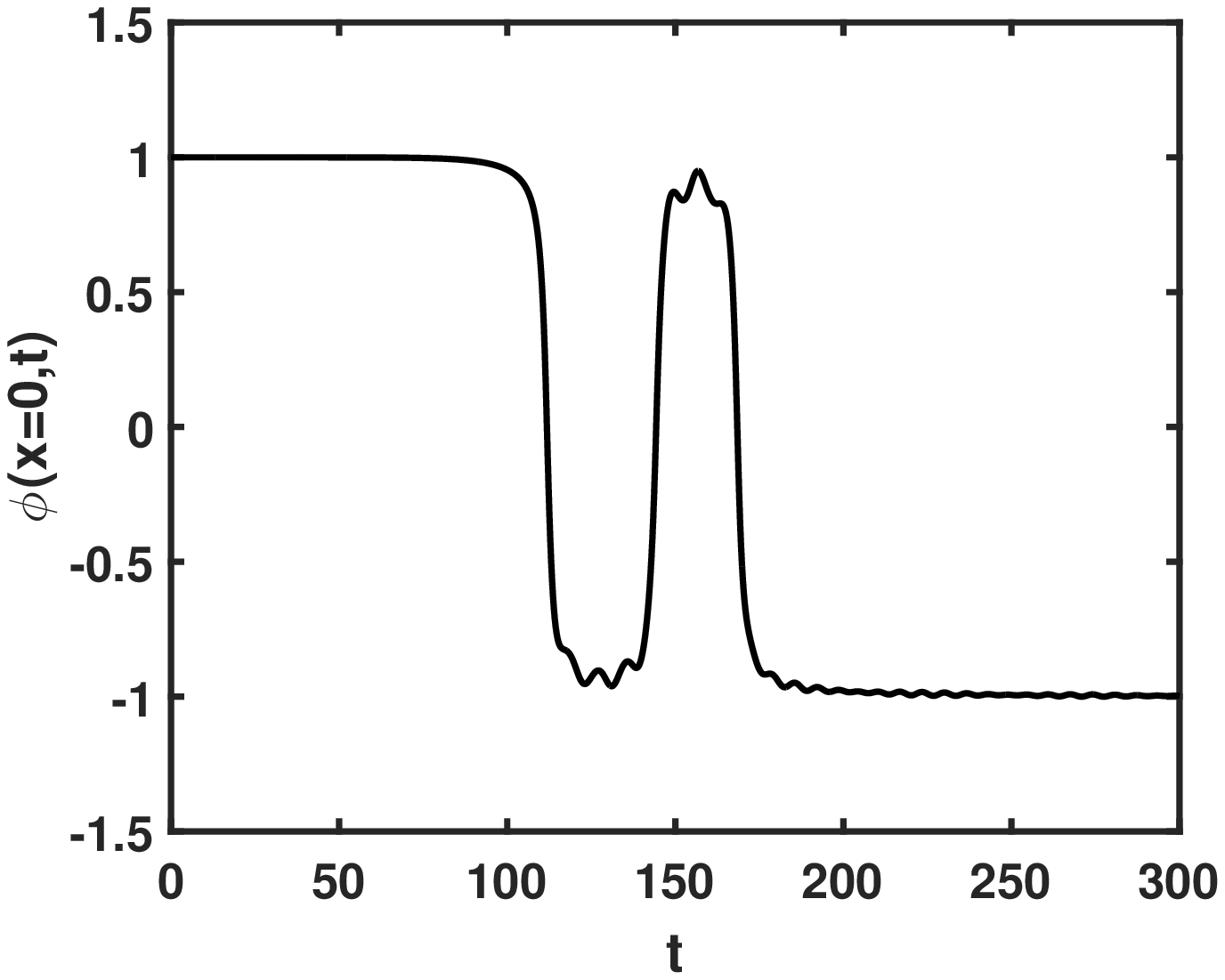}
\includegraphics[{angle=0,width=5cm, height=5cm}]{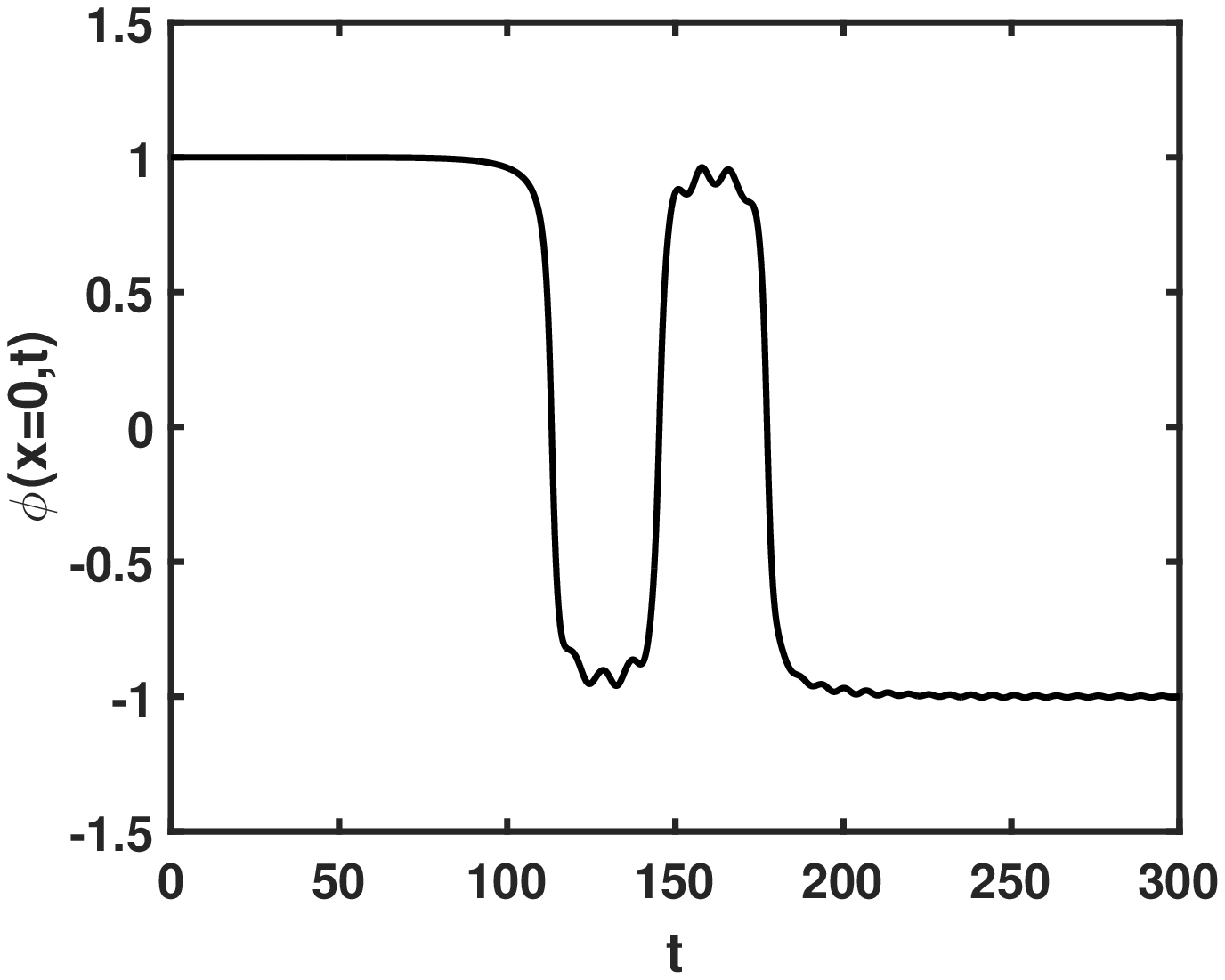}
\includegraphics[{angle=0,width=5cm, height=5cm}]{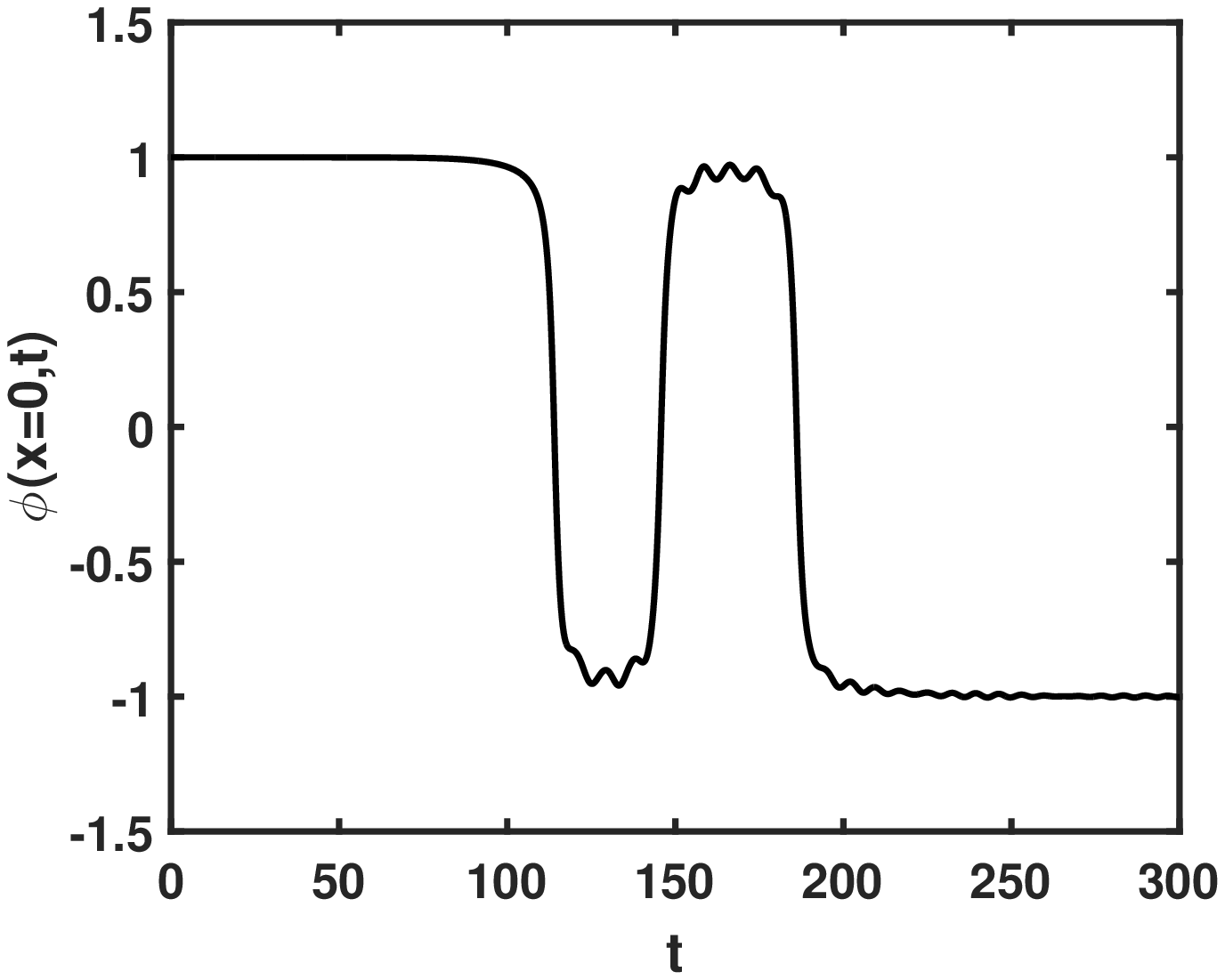}
\caption{Kink-antikink collisions: scalar field at the center of mass $\phi(x = 0, t)$ versus time for (a) $v=0.11069$, (b) $v=0.10918$ and (c) $v=0.10839$.}
\label{bounce_4}
\end{figure}

Fig. \ref{KA}b shows that close to the one-bounce windows we found another structure of thinner windows. These thinner windows appear for larger velocities, reducing progressively their thickness and accumulating around the maximum velocity of a given one-bounce window. An example of the peculiar structure of such collisions is depicted in Figs. \ref{bounce_4}a-c. Initially in the vacuum $\phi=1$ and after one bounce with a number $N$ of oscillations, the scalar field at the center of mass has another number $n$ of oscillations before tunneling to the other vacuum. The pair $(N,n)$ characterize the $n^{th}$ thinner window near to the $(N-1)^{th}$ one-bounce window. For example, Figs. \ref{bounce_4}a-c are characterized, respectively, by the pairs $(2,1)$, $(2,2)$, $(2,3)$. This means collisions corresponding to the first, second and third thinner windows, near the first one-bounce window.


\section { Conclusions  }


In this work we have investigated  antikink-kink  and kink-antikink  in a hybrid model. The model is similar to the $\phi^6$ model, in the sense that it engenders two distinct topological sectors. However, in each topological sector, there is a linear map between the hybrid model and the $\lambda\phi^4$ model. However, since the linear transformation is not the same for both topological sectors, it is not identical to the $\lambda\phi^4$ model. In each one of the two topological sectors the potential is symmetric around the local minima. The equation of motion has symmetric static kink and antikink solutions, and the stability analysis for kink and antikink result in a translational and a vibrational mode. Without loosing generality, we worked with initial configurations in the topological sector connecting vacua $\phi=0$ and $\phi=1$.

Our numerical investigation of antikink-kink scattering showed that, despite the models have different scalar field profiles, their structure of two- and three-bounce windows are related. Inside each window the structure of a particular scattering can be explained by the CSW mechanism, with the vibrational and continuum modes explaining qualitative and quantitatively the results. This corroborates the statement that there is no crossing to the other topological sector and that there is indeed a mapping between the hybrid model and the $\lambda\phi^4$ model for $\phi>0$.

\begin{figure}
\includegraphics[{angle=0,width=4cm,height=4cm}]{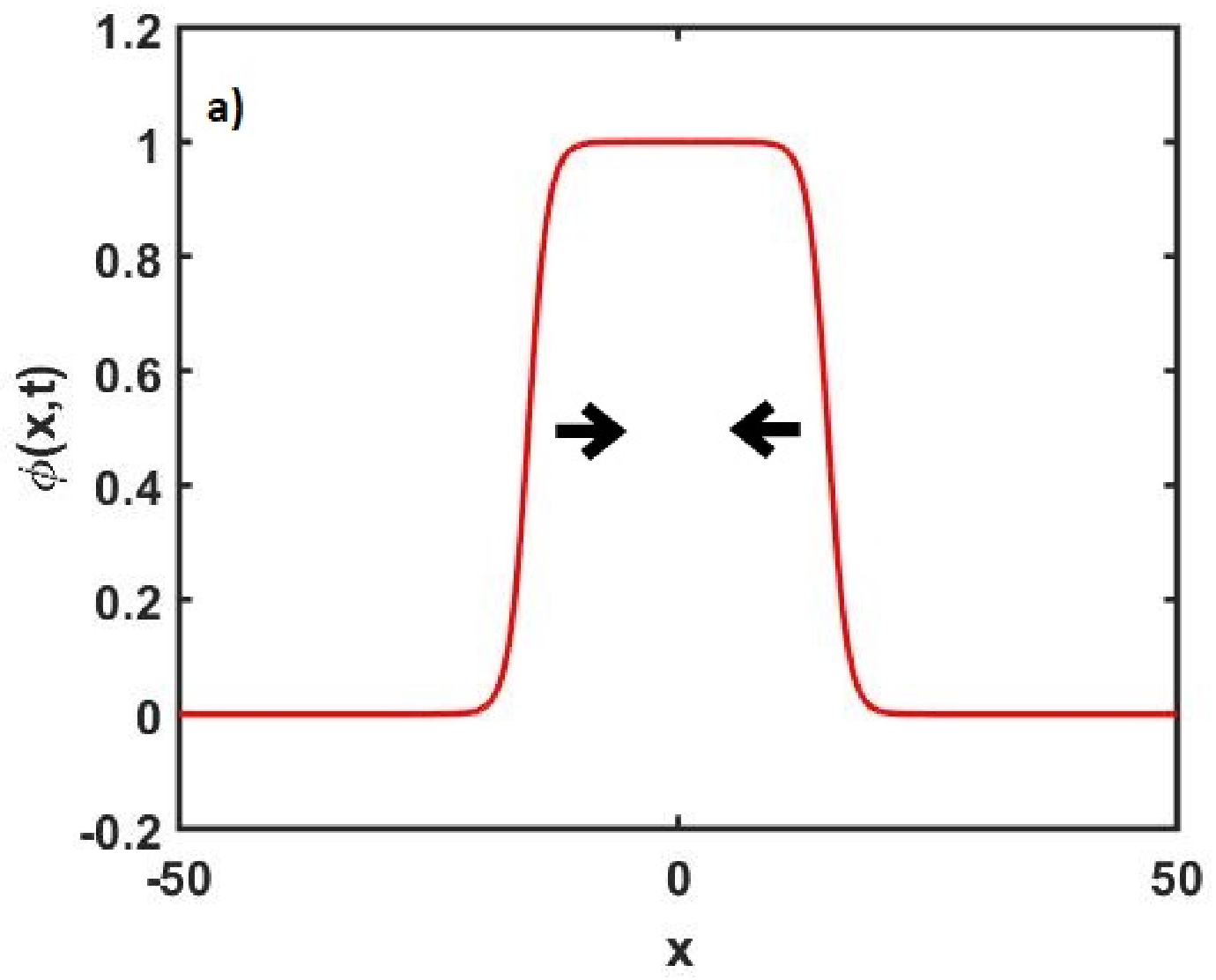}
\includegraphics[{angle=0,width=4cm,height=4cm}]{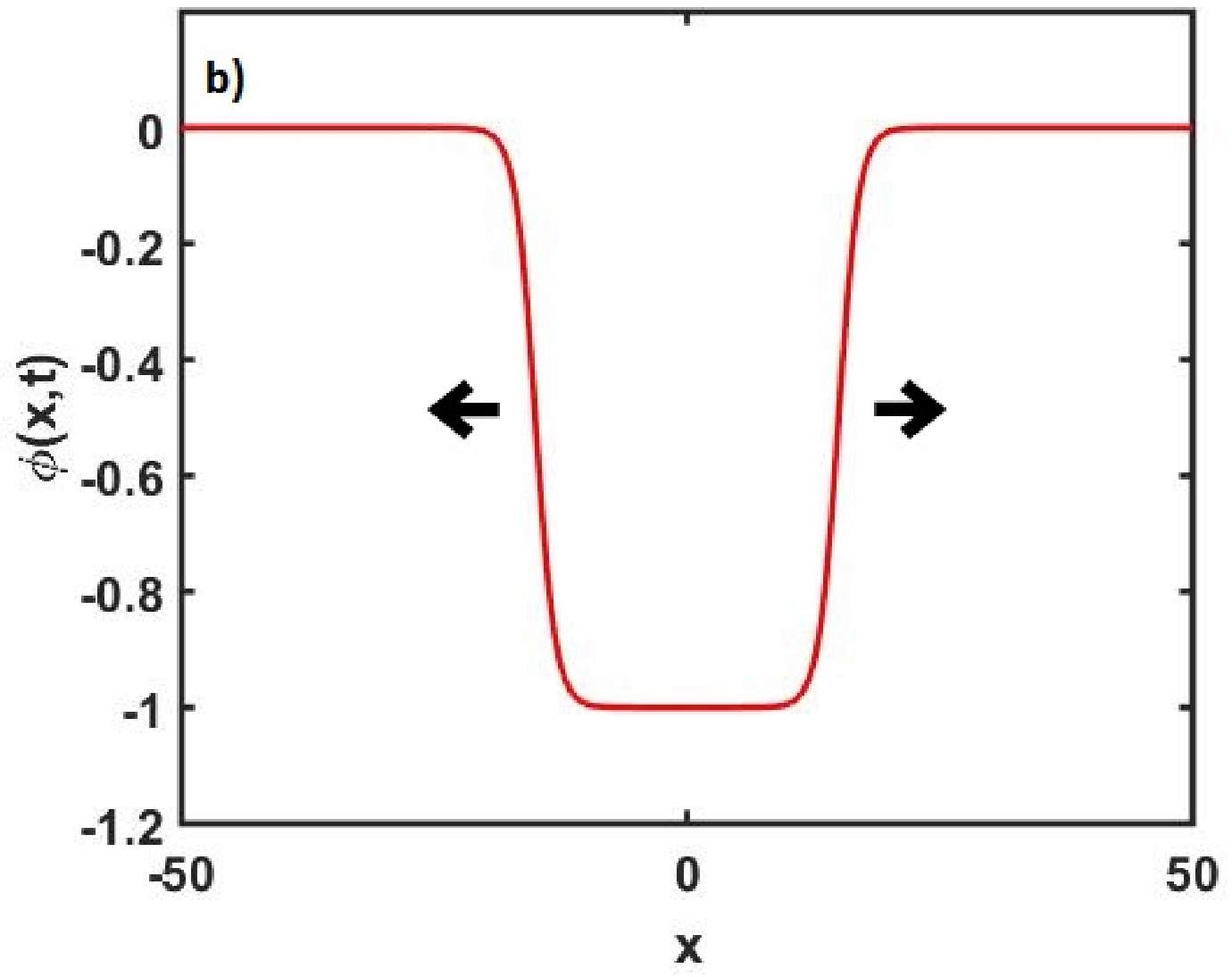}
\includegraphics[{angle=0,width=4cm,height=4cm}]{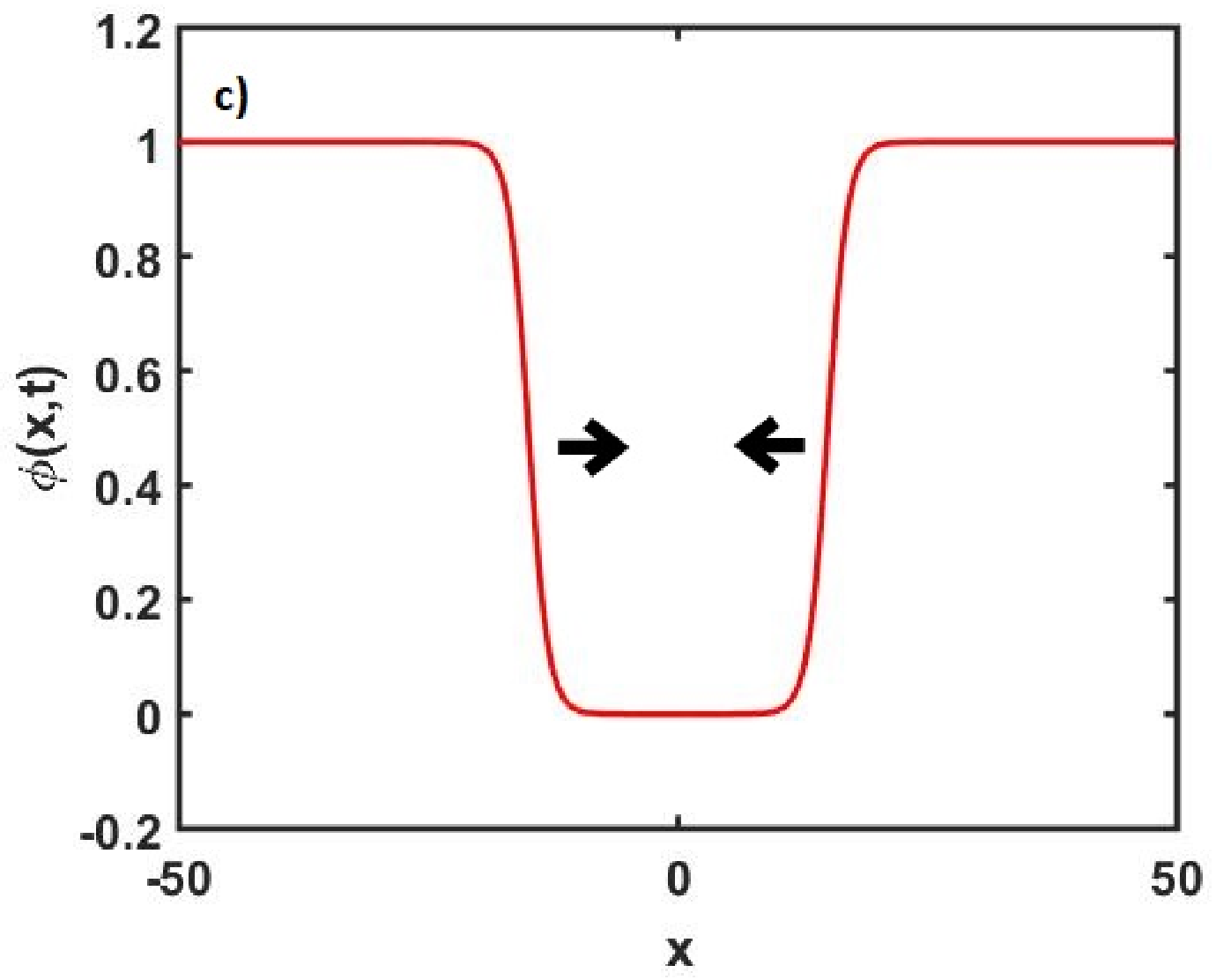}
\includegraphics[{angle=0,width=4cm,height=4cm}]{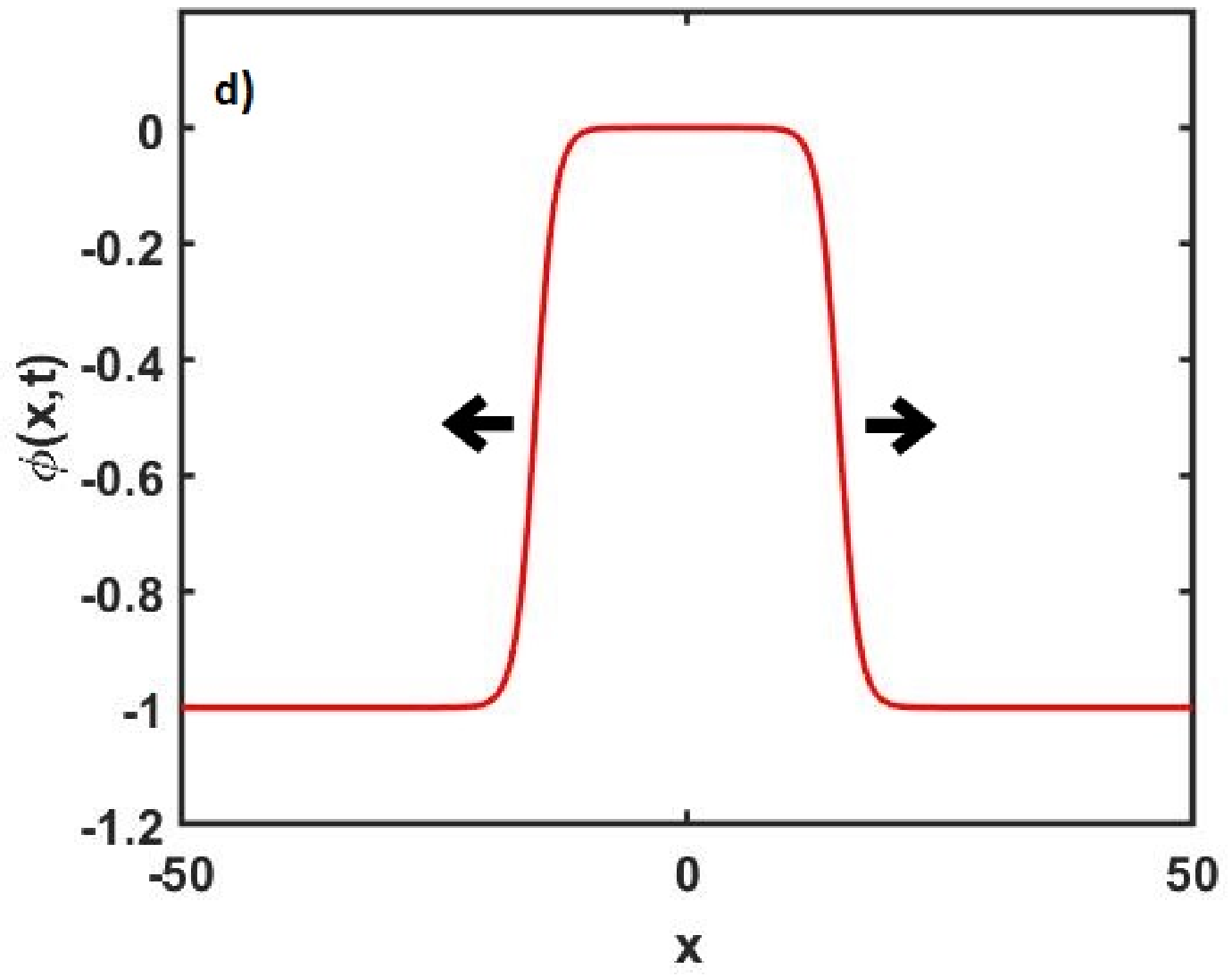}
\caption{Kink-antikink a) before and b) after collision resulting in a change of topological sector, as $K\bar K\to \bar K K$. An impossible antikink-kink c) before and d) after collision with a change of topological sector, as $\bar K K\to K \bar K $.}
\label{topol}
\end{figure}

The structure of kink-antikink scatering is richer. Indeed, for large velocities the pair has the possibility of changing to the other topological sector. This means that a linear map to the $\lambda\phi^4$ model for the whole process is not possible, resulting in a different scattering structure not reported before in the literature. The critical velocity separating one-bounce and bion states are smaller than the observed for antikink-kink scattering. This means that a transition from one topological sector to the other is favored in comparison to bion states, where the pair oscillated around the same topological sector. We showed that the one-bounce windows satisfy an adapted CSW mechanism. Also there are windows substructures characterized by a change of topological sector. This could be represented as $K\bar K\to \bar K K$, and the solutions change from the topological sector $(0,1)$ to the sector $(-1,0)$ as described in Figs. \ref{topol}a-b. A similar process for an antikink-kink, $\bar K K\to K \bar K $ would mean a changing from the sector $(1,0)$ to the sector $(-1,0)$ as described in Figs. \ref{topol}c-d. This however is not possible since it would demand an infinite amount of energy to change the vacuum in an infinite lenght interval.

The changing of the parameters $\lambda$ and $m$ do not modify the structure of bounce windows. Their effect can be seen in the scalar field at the center of mass in two ways: ii) since the vibrational frequency scales with $m$, larger values of $m$ shorten the time interval between bounces; ii) they fix the nonzero vacua as $\pm m/\sqrt{\lambda}$. The parameters also can have determinat influence on the pattern of emitted radiation of bion states, not studied in this work. Indeed, the energy density and the non-null vacua of the potential grows with the decreasing of the parameter $\lambda$, resulting in a larger rate of emitted radiation. This effect was studied recently in a model with a false vacuum that differs only slightly from the $\phi^4$ model \cite{adsinoav}. The increasing of parameter $m$ grows the frequency $\omega_k$ of the continuum modes - according to Eq. (\ref{omegak}) - which form the natural basis for the description of scalar radiation.  

If we had chosen initial configurations in the other topological sector $(0,-1)$/$(-1,0)$, the symmetry of the potential guarantees the exchange of behaviors for $K \bar K$ and $\bar K K$ collisions when compared the results described above. This would mean two-bounce windows with an structure identical to the $\phi^4$ model for kink-antikink collisions and the new behavior described above now for antikink-kink collisions. We would have the possibility of $\bar K K\to K \bar K $ with the changing of topological sector, but not for $K\bar K\to \bar K K$.

We also investigated kink-kink scattering. To solve the equation of motion we used the following initial conditions: $\phi(x,0)=\phi_K^{(0,\phi_v)}(x+x_0,v,0)+\phi_K^{(0,\phi_v)}(x-x_0,-v,0)-m/\sqrt\lambda$ and $\dot\phi(x,0)=\dot\phi_K^{(0,\phi_v)}(x+x_0,v,0)+\dot\phi_K^{(0,\phi_v)}(x-x_0,-v,0)$, with $x_0$ sufficiently large for strongly reduce the overlapping between the two kinks (for instance, for $\lambda=m^2=2$, the value  $x_0=12$ is enough). That is, the initial profile is $\phi(x,0)=\phi_K^{(-\phi_v,0)}(x+x_0,v,0)+\phi_K^{(0,\phi_v)}(x-x_0,-v,0)$, meaning that the kinks come from two different topological sectors. The numerical method was the same used described in the last section for antikink-kink and kink-antikink scattering. The kink-kink pair has a repulsive interaction, so cannot form a bound state. Our numerical anaysis showed that  after the collision the scalar field maintains the shape of the initial profile. For lower velocities the pair does not even touch, whereas for higher velocities there is one-bounce collision. The same applies for antikink-antikink collisions.

Starting from the two-vacua $\lambda\phi^4$ model, the linear mappings considered here are the most general possibles to construct a model with three vacua. An analogous procedure of linear mappings can be applied for more general models with an odd number $2n$ of vacua for generating the hybrid models with $2n+1$ vacua. The effects of the reflexing symmetry of the construction on kink scattering described here are expected to follow a similar pattern for such models.


\section{Acknowledgements}
A.R.G, K.Z.N and F.C.S. thank FAPEMA – Fundação de Amparo
à Pesquisa e ao Desenvolvimento do Maranhão through
grants PRONEX 01452/14, PRONEM 01852/14, Universal 01061/17,
01191/16, 01332/17 and 01441/18. A.R.G and D.B. thank CNPq
(brazilian agency) through grants 437923/2018-5, 311501/2018-4
and 306614/2014-6 for financial support. This study was financed
in part by the Coordenação de Aperfeiçoamento de Pessoal de
Nível Superior – Brasil (CAPES) – Finance Code 001.

\end{document}